\documentclass[showpacs,preprintnumbers,amsmath,amssymb]{revtex4}
\usepackage{graphicx}
\usepackage{epsfig}
\usepackage{dcolumn}
\usepackage{bm}
\usepackage{longtable}
\begin{document}

\title{\bf Study of the process $e^+e^- \to \pi^+\pi^-$ in the energy
           region $400<\sqrt[]{s}<1000$ MeV.}
\author{ M.N.Achasov}\email{achasov@inp.nsk.su}
\author{ K.I.Beloborodov} 
\author{A.V.Berdyugin}
\author{A.G.Bogdanchikov}
\author{A.V.Bozhenok} 
\author{A.D.Bukin}
\author{D.A.Bukin}
\author{T.V.Dimova} 
\author{V.P.Druzhinin} 
\author{V.B.Golubev} 
\author{I.A.Koop} 
\author{A.A.Korol} 
\author{S.V.Koshuba}
\author{A.P.Lysenko} 
\author{A.V.Otboev} 
\author{E.V.Pakhtusova} 
\author{S.I.Serednyakov} 
\author{Yu.M.Shatunov}
\author{V.A.Sidorov}
\author{Z.K.Silagadze} 
\author{A.N.Skrinsky}
\author{Yu.A.Tikhonov}
\author{A.V.Vasiljev}
\affiliation{ 
         Budker Institute of Nuclear Physics,  \\
         Siberian Branch of the Russian Academy of Sciences \\
         11 Lavrentyev,Novosibirsk,630090, Russia \\
	 Novosibirsk State University, \\
         630090, Novosibirsk, Russia}

\begin{abstract}
 The cross section of the process $e^+e^-\to \pi^+\pi^-$ was measured
 in the SND experiment at the VEPP-2M collider in the energy region 
 $400<\sqrt[]{s}<1000$ MeV. This measurement was based
 on about $12.4 \times 10^6$ selected collinear events, which include
 $7.4\times 10^6$ $e^+e^-\to e^+e^-$, $4.5\times 10^6$ $e^+e^-\to\pi^+\pi^-$
 and $0.5\times 10^6$ $e^+e^-\to\mu^+\mu^-$ selected events.
 The systematic  uncertainty of the cross section determination is 1.3 \%.
 The $\rho$-meson parameters were determined:  $m_\rho=774.9\pm 0.4\pm 0.5$ 
 MeV, $\Gamma_\rho=146.5\pm 0.8\pm 1.5$ MeV, 
 $\sigma(\rho\to\pi^+\pi^-)=1220\pm 7\pm 16$ nb as well as the parameters of 
 the $G$-parity suppressed decay $\omega\to\pi^+\pi^-$: 
 $\sigma(\omega\to\pi^+\pi^-)=29.9\pm 1.4\pm 1.0$ nb and 
 $\phi_{\rho\omega} = 113.5\pm 1.3\pm 1.7$ degree.
\end{abstract}


\maketitle

\section{Introduction}

 The cross section of the $e^+e^-\to\pi^+\pi^-$ process in the energy region
 $\sqrt[]{s}<1000$ MeV can be described within the vector meson dominance model
 (VDM) framework and is determined by the transitions $V\to\pi^+\pi^-$ of the 
 light vector
 mesons ($V=\rho,\omega,\rho^\prime,\rho^{\prime\prime}$) into the final 
 state. The main contribution in this energy region comes
 from the $\rho\to\pi^+\pi^-$ and from the G-parity violating 
 $\omega\to\pi^+\pi^-$ transitions. Studies of the
 $e^+e^-\to\pi^+\pi^-$ reaction allow us to determine the $\rho$ and
 $\omega$ meson parameters and provide information on the $G$-parity
 violation mechanism.

 At low energies the $e^+e^-\to\pi^+\pi^-$ cross section gives the 
 dominant contribution to the celebrated ratio
 $R(s)=\sigma(e^+e^-\to\mbox{hadrons})/\sigma(e^+e^-\to\mu^+\mu^-)$,
 which is used for calculation of the dispersion integrals. For example,
 for evaluation of the electromagnetic running coupling constant at the 
 $Z$-boson mass $\alpha_{em}(s=m_Z^2)$, or for determination of the hadronic 
 contribution  $a^{hadr}_\mu$
 to the anomalous magnetic moment of the muon, which nowadays is
 measured with very high accuracy $5\times 10^{-6}$ \cite{bnl1,bnl2}.

 Assuming conservation of the vector current (CVC) in the isospin symmetry
 limit,
 the spectral function of the  $\tau^\pm\to\pi^\pm\pi^0\nu_\tau$ decay can
 be related to the isovector part of the $e^+e^-\to\pi^+\pi^-$ cross section.
 The spectral function was determined with high precision in
 Ref.\cite{aleph,opal,cleo2}. The comparison of the  $e^+e^-\to\pi^+\pi^-$
 cross section with what follows from the spectral function provides 
 an accurate test of the CVC hypothesis.

 The process $e^+e^-\to\pi^+\pi^-$ in the energy region $\sqrt[]{s}<1000$ MeV
 was studied in several experiments \cite{augu,ausl,bena,quen,vas1,buki,vas2,
 vas3,kur1,kur2,spec,olya,kmd2,kloe} during more than 30 years. In present 
 work the results of the $e^+e^-\to\pi^+\pi^-$
 cross section measurement with SND detector at $390\le\sqrt{s}\le 980$ MeV
 are reported.
 
\section{Experiment}

 The SND detector \cite{sndnim} operated from 1995 to 2000 at the
 VEPP-2M \cite{vepp2} collider in the energy range $\sqrt[]{s}$ from 360 to
 1400 MeV. The detector contains several subsystems. The tracking system
 includes two cylindrical drift chambers. The three-layer spherical
 electromagnetic calorimeter is based on NaI(Tl) crystals.
 The muon/veto system consists of plastic scintillation counters and two
 layers of streamer tubes. The calorimeter energy and angular resolutions
 depend on the photon energy as
 $\sigma_E/E(\%) = {4.2\% / \sqrt[4]{E(\mbox{GeV})}}$ and
 $\sigma_{\phi,\theta} = {0.82^\circ / \sqrt[]{E(\mathrm{GeV})}} \oplus
 0.63^\circ$. The tracking system angular resolution is about $0.5^\circ$ and
 $2^\circ$ for azimuthal and polar angles respectively.
 
 In 1996 -- 2000 the SND detector collected data in the energy region
 $\sqrt[]{s}<980$ MeV with integrated luminosity about $10.0~\mbox{pb}^{-1}$.
 The beam energy was calculated from the magnetic field value in the bending
 magnets of the collider. The accuracy of the energy setting is about 0.1 
 MeV. The beam energy spread varies in the range from 0.06 MeV at
 $\sqrt[]{s}= 360$ MeV to 0.35 MeV at $\sqrt[]{s}=970$ MeV.

\section{Data Analysis}

 The cross section of the $e^+e^-\to\pi^+\pi^-$ process was measured in the
 following way. 
\begin{enumerate}
\item
 The collinear events $e^+e^-\to e^+e^-,\pi^+\pi^-,\mu^+\mu^-$ were selected;
\item
 The selected events were sorted into the two classes: $e^+e^-$ and  
 $\pi^+\pi^-,\mu^+\mu^-$ using the energy deposition in the calorimeter 
 layers;
\item
 The $e^+e^-\to e^+e^-$ events were used for integrated luminosity
 determination. The events of the $e^+e^-\to\mu^+\mu^-$ process were 
 subtracted according to the theoretical cross section, integrated
 luminosity and detection efficiency;
\item
 In order to determine the cross section of the 
 $e^+e^-\to\pi^+\pi^-$ process, the number of $e^+e^-\to\pi^+\pi^-$
 events in each energy point were normalized on the integrated luminosity
 and divided by the detection efficiency and radiative correction.
\end{enumerate}

 The detection efficiency was obtained from Monte Carlo (MC) simulation
 \cite{sndnim}. The MC simulation of SND is based on UNIMOD \cite{unimod}
 package. The SND geometrical model description comprises about 10000 
 distinct  volumes and
 includes details of the SND design. The  primary generated particles are
 tracked through the detector media taking into account the following
 effects: ionization losses, multiple scattering, 
 bremsstrahlung of electrons
 and positrons, Compton effect and Rayleigh scattering, $e^+e^-$ pair
 production by photons, photo-effect, unstable particles decays, interaction 
 of stopped particles, nuclear interaction
 of hadrons \cite{union,umnuc1,umnuc2}. 
 After that the signals produced in each detector element are simulated.
 The electronics noise, signals pile up, the actual time and amplitude 
 resolutions of the electronics channels and broken channels were taken into
 account during processing the Monte Carlo events to provide the adaptable 
 account of variable experimental conditions.

 The MC simulation of the processes $e^+e^-\to e^+e^-,\mu^+\mu^-,\pi^+\pi^-$
 was based on the formula obtained in the Ref.\cite{berkl,arbuzqed,arbuzhad}.
 The simulation of the process  $e^+e^-\to e^+e^-$ was performed with the 
 cut $30^\circ<\theta_{e^\pm}<150^\circ$ on the polar angles of the final 
 electron and positron. 
 
\begin{figure}
\epsfig{figure=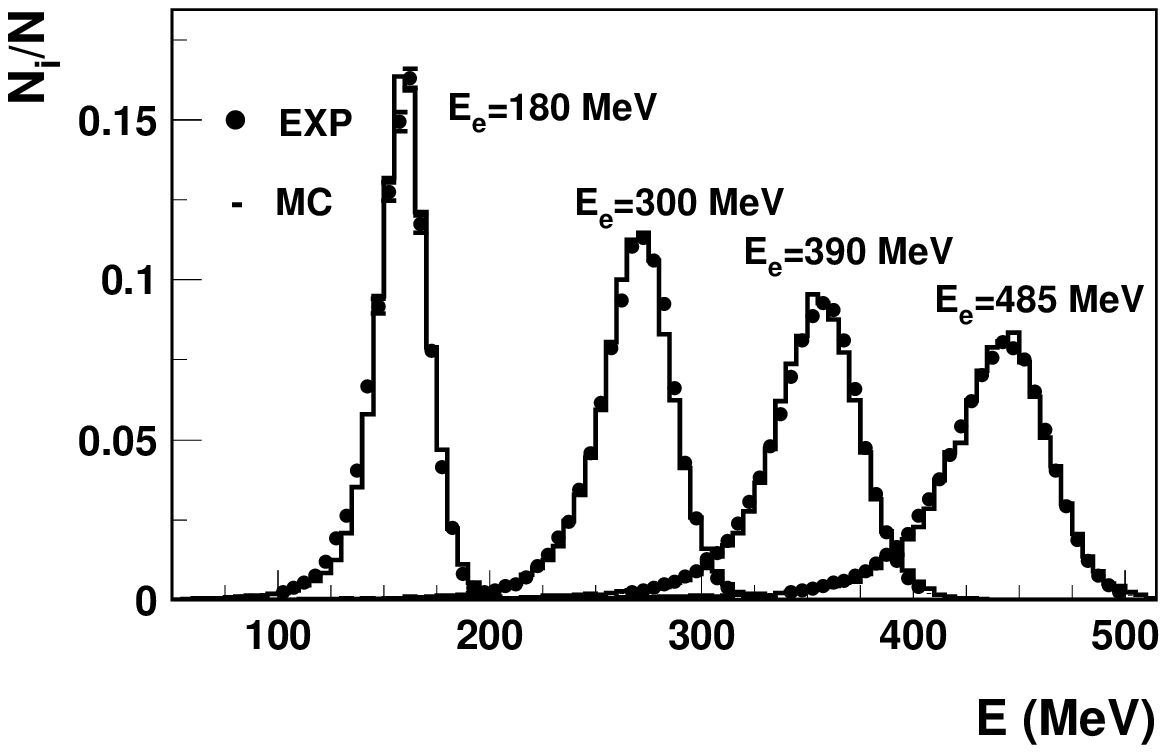,width=15.0cm}
\caption{Energy deposition spectra for electrons with the energies
         180, 300, 390 and 485 MeV in experiment (dots) and simulation
         (histogram).}
\label{enee}
\epsfig{figure=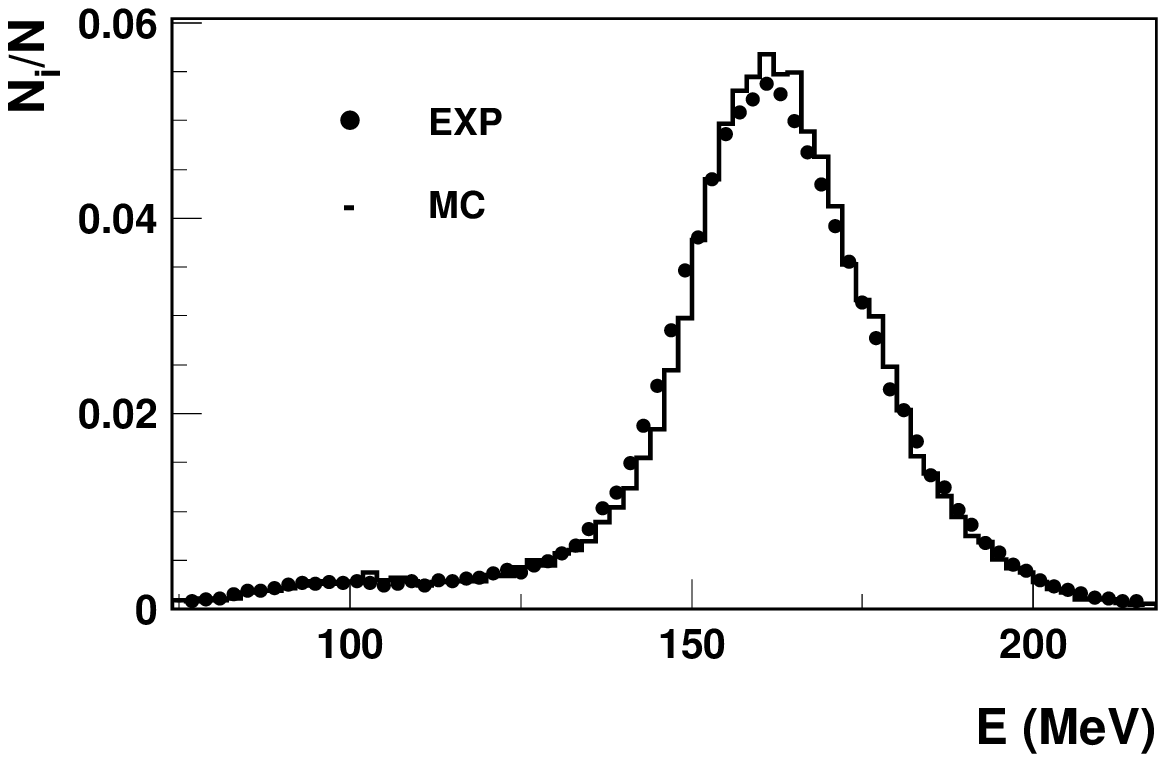,width=15.0cm}
\caption{Energy deposition spectra for the 500 MeV muons in experiment (dots)
         and simulation (histogram).}
\label{enmm}
\end{figure}

\begin{figure}
\epsfig{figure=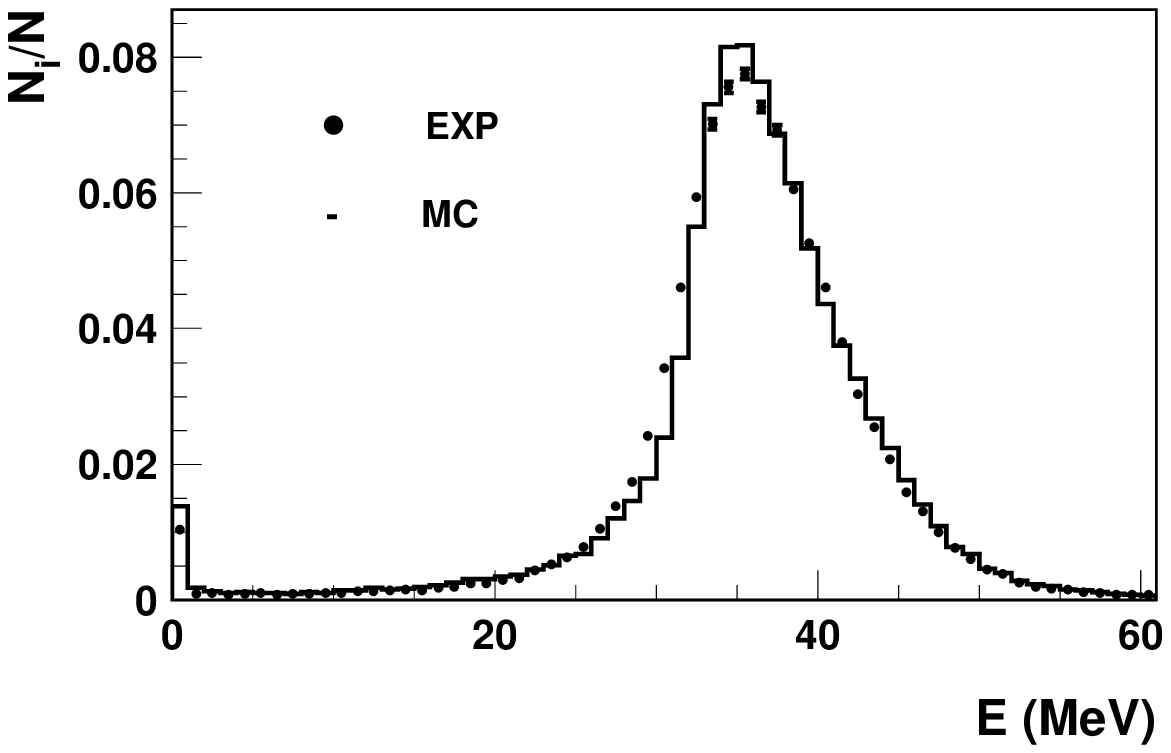,width=15.0cm}
\caption{The spectra of the ionization losses of the pions with energy 
         $E_\pi>360$ MeV in the first calorimeter layer. Dots -- experiment, 
	 histogram -- simulation.}
\label{enppi}
\epsfig{figure=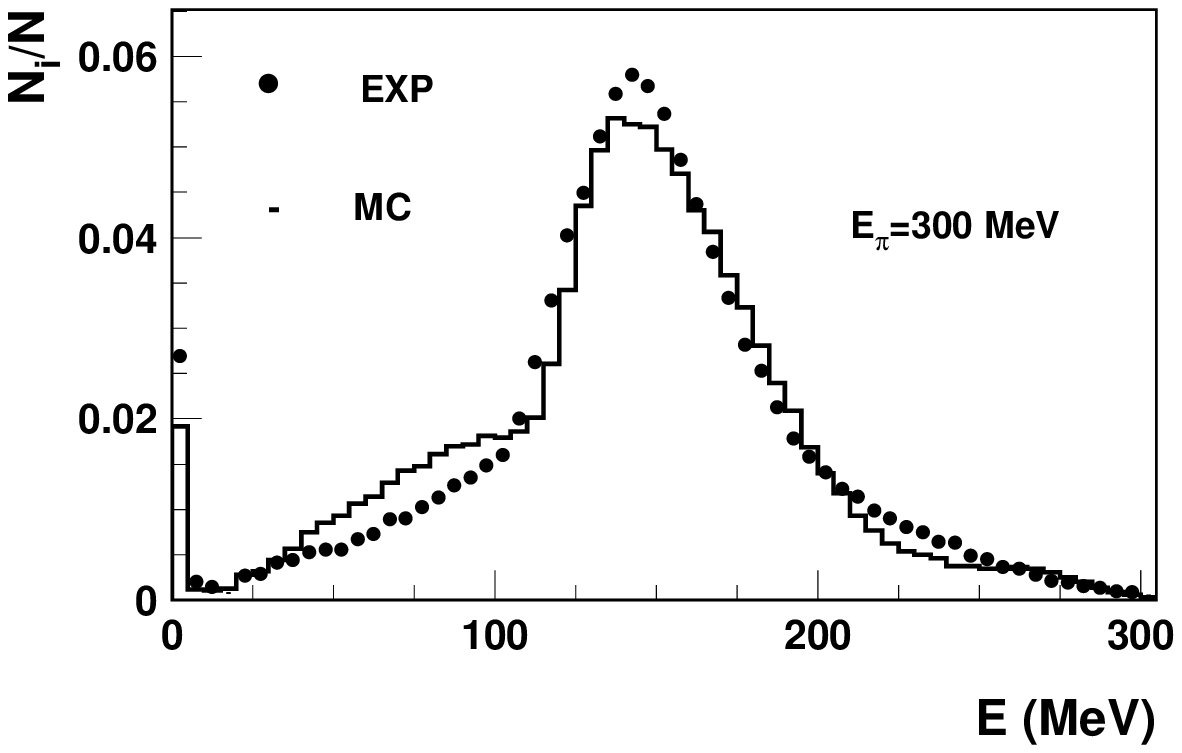,width=15.0cm}
\caption{Energy deposition spectra of the pions with the energy $E_\pi=300$
         MeV. Dots -- experiment, histogram -- simulation.}
\label{enpi300}
\end{figure}

 The $e^+e^-\to e^+e^-$, $\mu^+\mu^-$ and $\pi^+\pi^-$ events are 
 differed by energy deposition in the calorimeter. In  $e^+e^-\to e^+e^-$ 
 events the electrons produce the electromagnetic shower with the most 
 probable 
 energy losses about 0.92 of the initial particle energy.
 The distributions of the energy deposition of the electrons with the 
 different energies are shown in Fig.\ref{enee}. The experimental and simulated
 spectra are in good agreement. Muons lose their energy by ionization 
 of the calorimeter material through which they pass and
 their energy deposition spectra are well modeled in simulation 
 (Fig.\ref{enmm}). The similar ionization losses are experienced by charged 
 pions and this part of the charged pion energy deposition is well described
 by simulation (Fig.\ref{enppi}). But pions lose their energy also due to
 nuclear interactions which is not so accurately reproduced in simulation.
 This leads to some difference in energy deposition spectra in 
 experiment and simulation for charged pions (Fig.\ref{enpi300}). 

\begin{figure}
\epsfig{figure=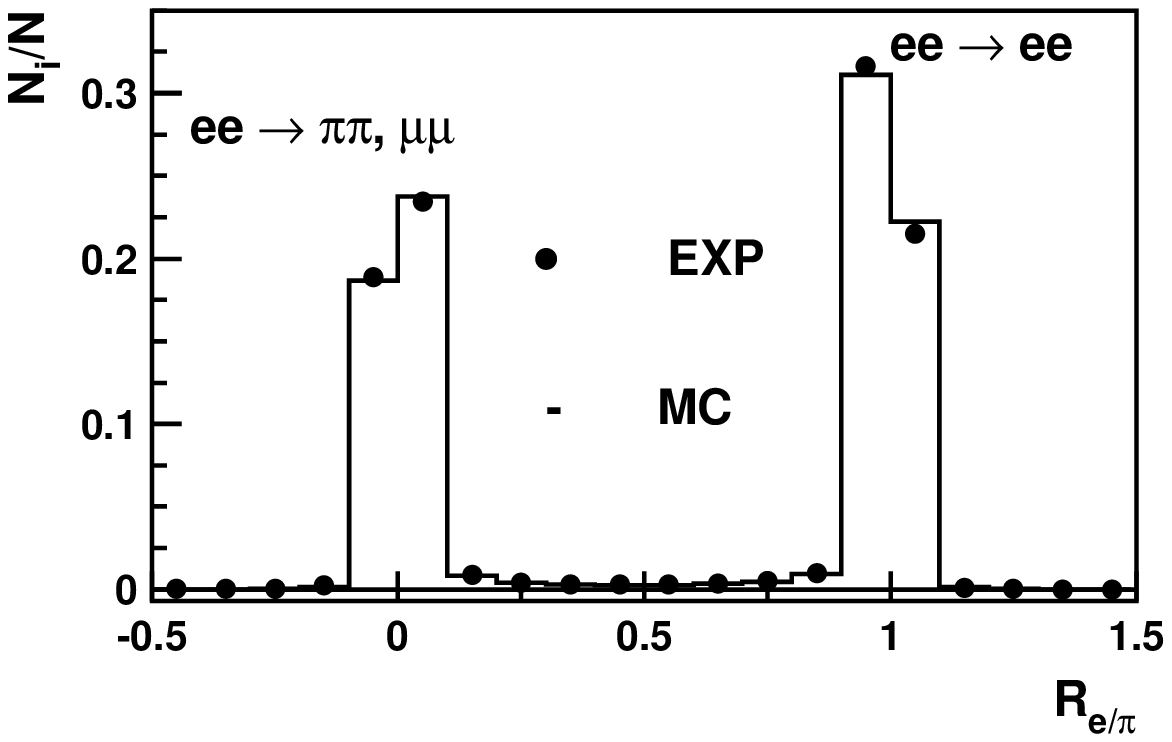,width=15.0cm}
\caption{The $e/\pi$ discrimination parameter distribution for all collinear
         events in the energy region $\sqrt{s}$ from 880 to 630  MeV.
	 Dots -- experiment, histogram -- simulation.}
\label{mlp2}
\end{figure}

 The discrimination between electrons and pions in the SND detector is
 based on difference in longitudinal energy deposition profiles (deposition
 in calorimeter layers) for these particles. To use in the most complete way
 the correlations between energy depositions in the calorimeter layers,
 the corresponding separation parameter was based
 on the neural network approach \cite{neural1}. For each energy
 point the neural network -- multilayer perceptron was constructed.
 The network had input layer consisting of 7 neurons, two hidden layers
 with 20 neurons each and the output layer with one neuron.
 As the input data the network used the energy depositions of the particles in
 calorimeter layers and the polar angle of one of the particles. The output 
 signal $R_{e/\pi}$ is a number in the interval from -0.5 to 1.5. The network 
 was trained by 
 using simulated $e^+e^-\to\pi^+\pi^-$ and $e^+e^-\to e^+e^-$ events.
 The distribution of the discrimination parameter $R_{e/\pi}$ is shown in
 Fig.\ref{mlp2}. The $e^+e^-\to e^+e^-$ events are located in the region
 $R_{e/\pi}>0.5$, while $e^+e^-\to\pi^+\pi^-,\mu^+\mu^-$ events at 
 $R_{e/\pi}<0.5$.


\subsection{Selection criteria}

 During the experimental runs, the first-level trigger \cite{sndnim} selects
 events with one or more tracks in tracking system and with two clusters in
 calorimeter with the spatial angle between the clusters more than
 $100^\circ$. The threshold on energy deposition in cluster was equal
 to 25 MeV. The threshold on the total energy deposition  in the calorimeter
 was set equal to 140 MeV in the energy region $\sqrt{s}\ge 850$ MeV,
 and to 100 MeV, or was absent at all, below 850 MeV.  
 During processing of the experimental data the event reconstruction is 
 performed \cite{sndnim,phi98}. For further analysis, events containing two
 charged particles with $|z| < 10$ cm and $r < 1$ cm were selected. Here $z$
 is the coordinate of the charged particle production point along the beam 
 axis (the longitudinal size of the interaction region depends on beam 
 energy and varies from 1.5 to 2.5 cm);  $r$ is the distance between the 
 charged particle track and the beam axis in the $r-\phi$ plane. 
 The polar angles of the charged particles were bounded by the criterion:
 $55^\circ<\theta<125^\circ$ and the energy deposition of each of them was
 required to be greater than 50 MeV. The following cuts on the acollinearity
 angles in the azimuthal and polar planes were applied:
 $|\Delta\phi|<10^\circ$ and $|\Delta\theta|<10^\circ$. In the event sample
 selected under these conditions one has the $e^+e^-\to e^+e^-$, 
 $\pi^+\pi^-$, $\mu^+\mu^-$ events, cosmic muons background and
 a small contribution from the $e^+e^-\to\pi^+\pi^-\pi^0$ reaction at 
 $\sqrt{s}\simeq m_\omega$. The muon system $veto$ was 
 used for suppression of the cosmic muon background ($veto=0$). 

\subsection{The background from the cosmic muons and from the 
$e^+e^-\to\pi^+\pi^-\pi^0$ process.}

\begin{figure}
\epsfig{figure=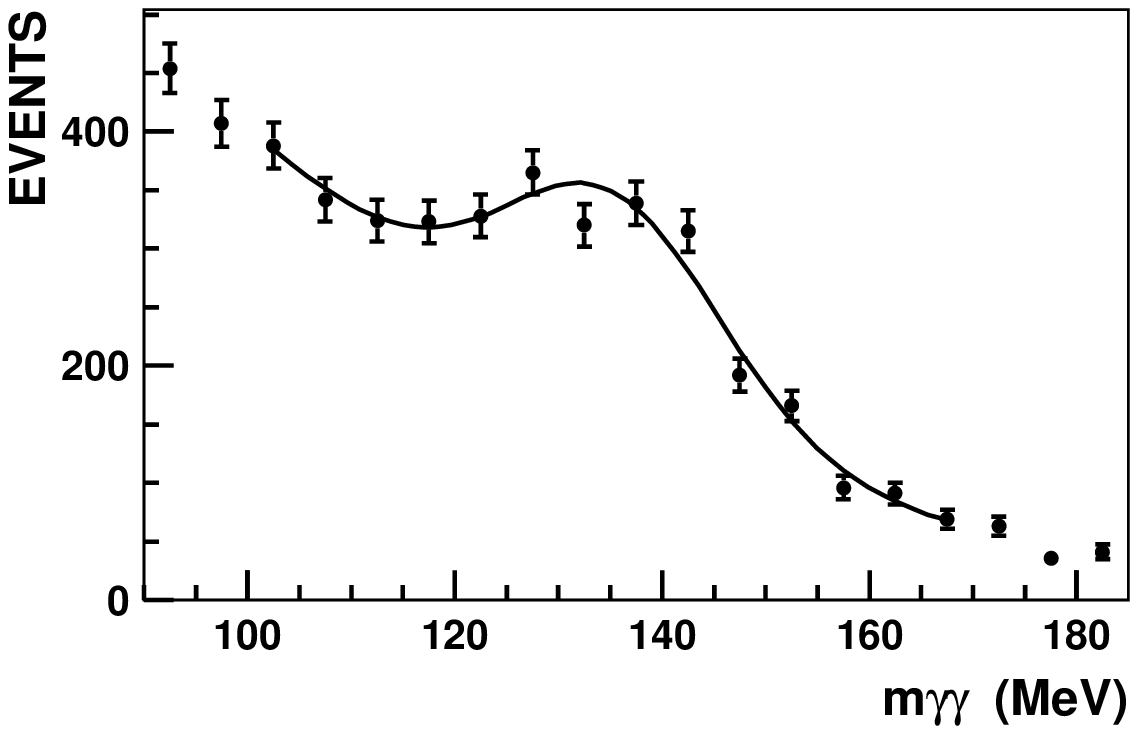,width=15.0cm}
\caption{Two-photon invariant mass $m_{\gamma\gamma}$ distribution at
         $\sqrt{s}\simeq m_\omega$.}
\label{mpi0}
\epsfig{figure=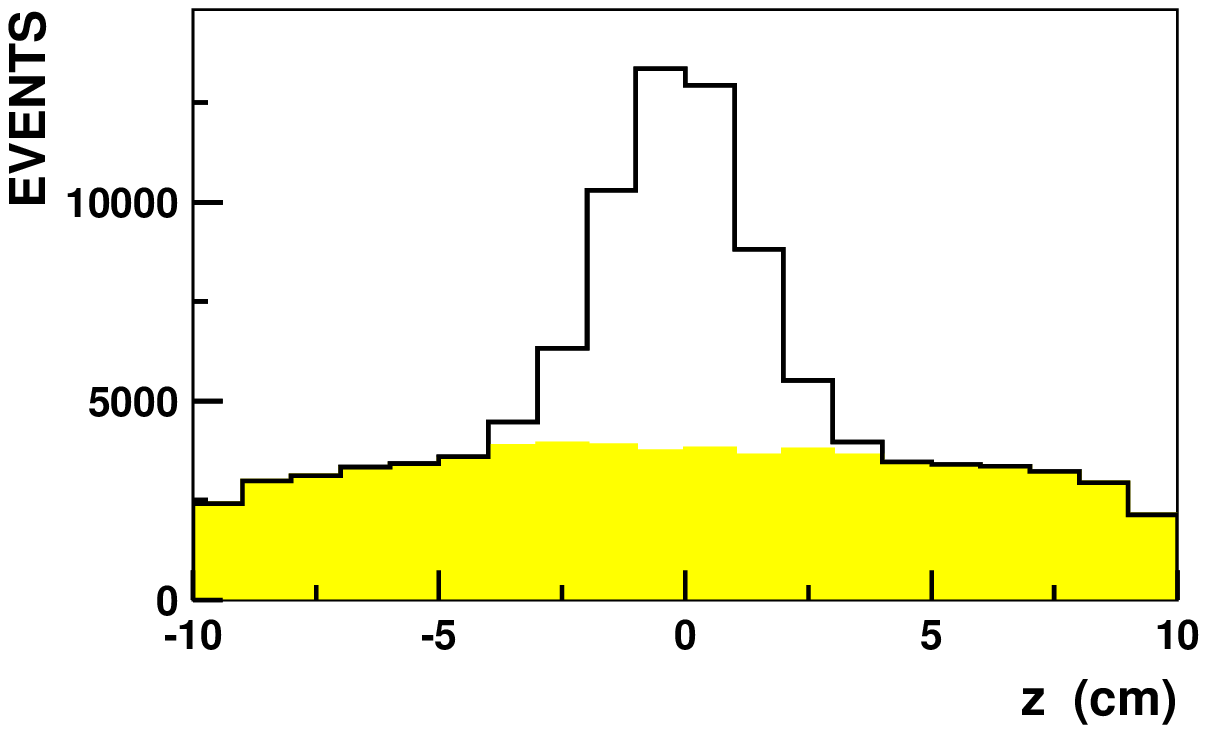,width=15.0cm}
\caption{The distribution of the $z$ coordinate of the charged particle
         production point along the beam axis for collinear events at
	 $\sqrt{s}=180$ MeV. Histogram -- all events, dashed distribution
	 -- events with muon system $veto$ ($veto=1$).}
\label{actz}
\end{figure}
 The number of background events from the $e^+e^-\to\pi^+\pi^-\pi^0$ process 
 was estimated in the following way:
\begin{eqnarray}
\label{bg}
 N_{3\pi}({s}) = \sigma_{3\pi}({s}) \epsilon_{3\pi}({s})  IL({s}),
\end{eqnarray}
 where $\sigma_{3\pi}({s})$ is the cross section of the
 $e^+e^-\to\pi^+\pi^-\pi^0$ process with the radiative corrections taken
 into account, $IL({s})$ is the integrated luminosity, $\epsilon_{3\pi}({s})$
 is the detection probability for the background process obtained from the
 simulation under the selection criteria described above. The values of
 $\sigma_{3\pi}({s})$ were taken from the SND measurements \cite{pi3omeg}.
 Although $\sigma_{3\pi}(m_\omega)\approx 1300$ nb, the $e^+e^-\to 3\pi$
 process contribution to the total number of the collinear events at the
 $\omega$ resonance peak is less than 0.3 \%.
 The leading role in the suppression of this background was played by the cuts
 on the acollinearity angles $\Delta\theta$ and $\Delta\phi$.
 In order to check the estimation (\ref{bg}), the events containing two and
 more photons with energy depositions more than 200 MeV were considered. 
 The constraint on the photons energy deposition greatly suppresses 
 not the $e^+e^-\to 3\pi$ events, as a result of the fact that our 
 selection criteria select the 
 $e^+e^-\to 3\pi$ events with collinear charged pions and therefore
 the neutral pion in this events has relatively low energy.  
 In order to obtain $e^+e^-\to 3\pi$ events number $n_{3\pi}$, 
 the invariant mass spectrum $m_{\gamma\gamma}$ (Fig.\ref{mpi0}) was
 fitted by the sum of Gaussian and the second order polynomial:
 $G(m_{\gamma\gamma})\times n_{3\pi}+P_2(m_{\gamma\gamma})\times(n-n_{3\pi})$.
 The value of $n_{3\pi}$ agrees with events number calculated according
 to (\ref{bg}).
 
 The cosmic muon background was suppressed by the muon/veto system. 
 The $z$ coordinate distribution for the  charged particle production point
 along the beam axis is shown in Fig.\ref{actz} for collinear events.
 The $e^+e^-$ annihilation events have the Gaussian distribution peaked
 at $z=0$, while the cosmic background distribution is nearly uniform
 and clearly extends outside the peak. As the Fig.\ref{actz} shows, the
 muon system $veto$ ($veto=1$) separates cosmic muons from the $e^+e^-$
 annihilation events. The residual events number of the cosmic muon 
 background was estimated from the following formula:
\begin{eqnarray}
 N_\mu=\nu_\mu \times T.
\end{eqnarray}
 Here $\nu_\mu\simeq 1.3\times 10^{-3}$ Hz is the frequency of cosmic 
 background registration under the applied selection criteria, 
 $T$ is the time of data taking. The value of $\nu_\mu$ was obtained by using
 data collected in special runs without beams in collider. The first-level 
 trigger counting rate in these runs was 2 Hz.  
 The contribution of the cosmic background to the total number of selected 
 collinear events depends on energy $\sqrt[]{s}$ and varies from 0.1 \%
 to 1 \%.

 The $e^+e^-\to\pi^+\pi^-\pi^0$ events are concentrated in the $R_{e/\pi}$
 discrimination parameter region $R_{e/\pi}<0.5$. The cosmic background
 events at the energies $\sqrt{s}>600$ also fall in the area $R_{e/\pi}<0.5$,
 because the energy deposition of the cosmic muons is much lower
 than the energy deposition in the $e^+e^-\to e^+e^-$ events. For the lower
 center of mass energies the cosmic background moves to the area 
 $R_{e/\pi}>0.5$, because in this case the energy depositions are close.

\subsection{Detection efficiency}

\begin{figure}
\begin{center}
\epsfig{figure=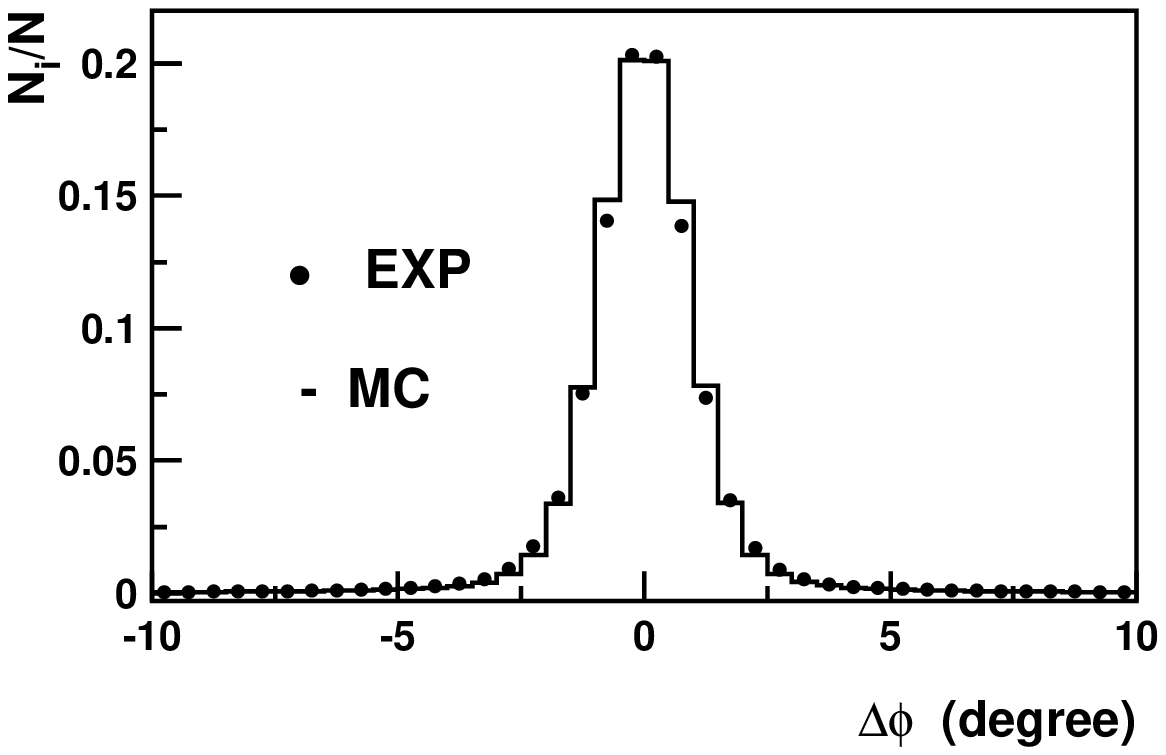,width=15.0cm}
\caption{The $\Delta\phi$ distribution of the $e^+e^-\to e^+e^-$
         events. Dots -- experiment, histogram -- simulation.}
\label{dphiee}
\epsfig{figure=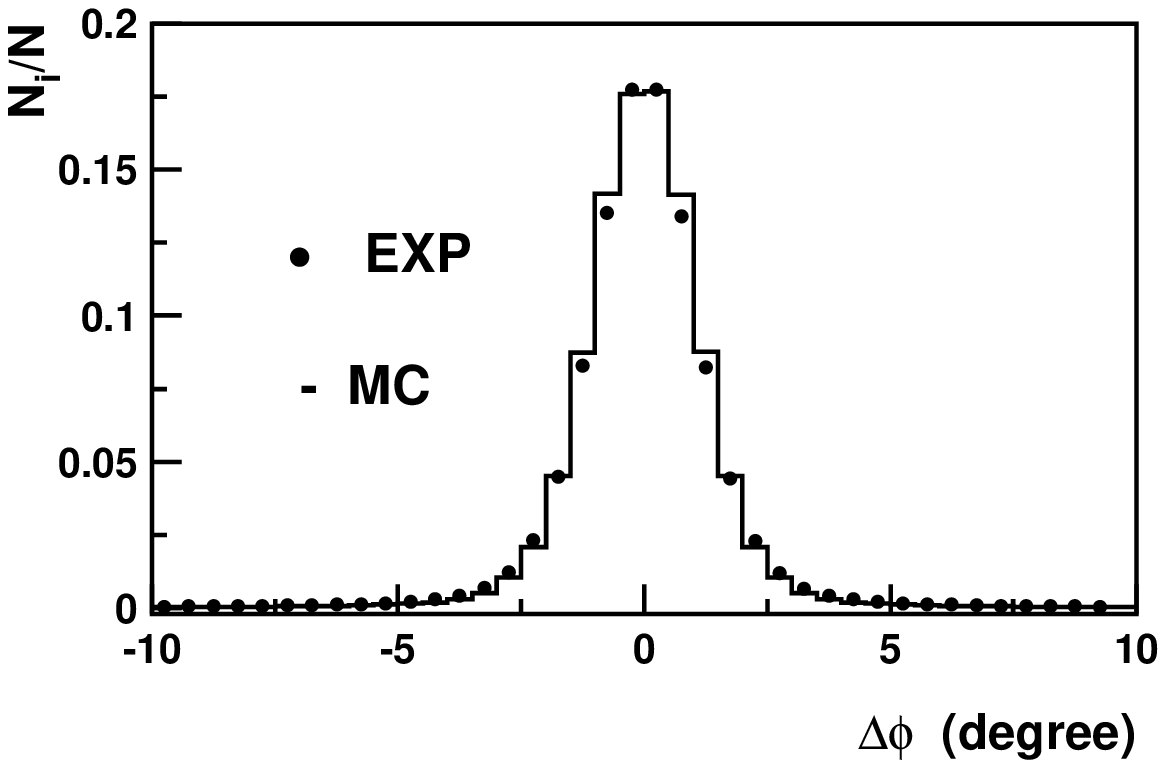,width=15.0cm}
\caption{The $\Delta\phi$ distribution of the $e^+e^-\to\pi^+\pi^-$
         events. Dots -- experiment, histogram -- simulation.}
\label{dphipp}
\end{center}
\end{figure}

\begin{figure}
\begin{center}
\epsfig{figure=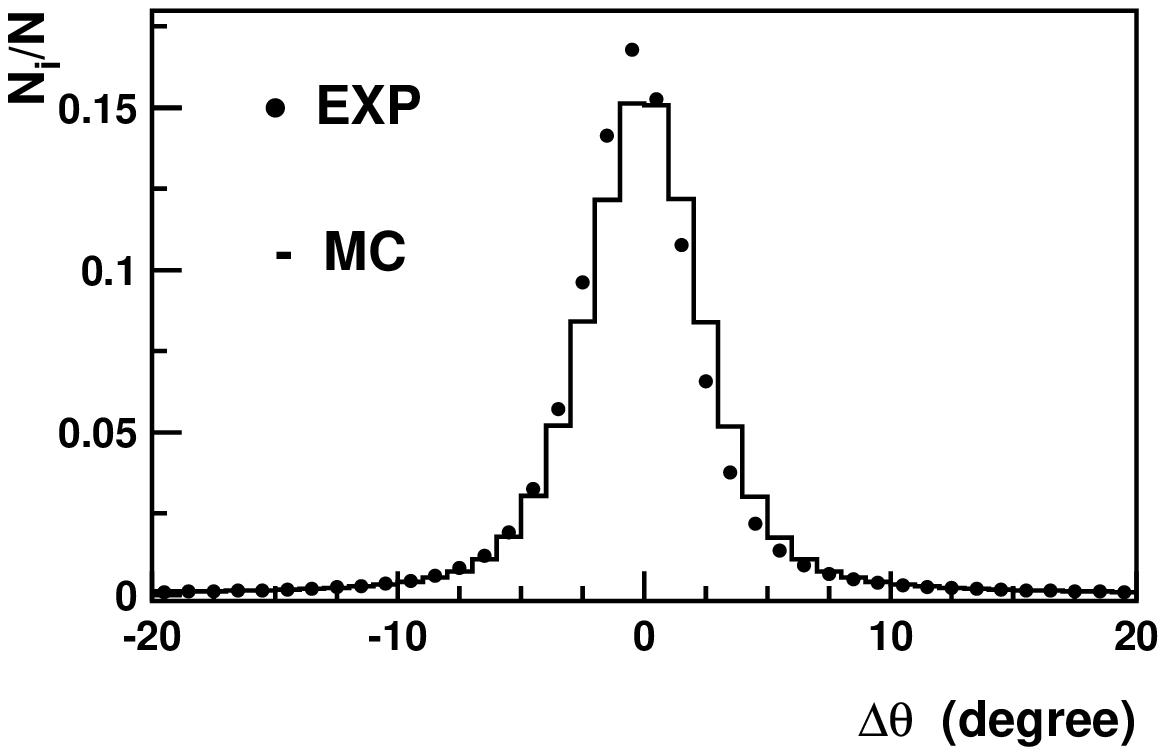,width=15.0cm}
\caption{The $\Delta\theta$ distribution of the $e^+e^-\to e^+e^-$
         events. Dots -- experiment, histogram -- simulation.}
\label{dtetee}
\epsfig{figure=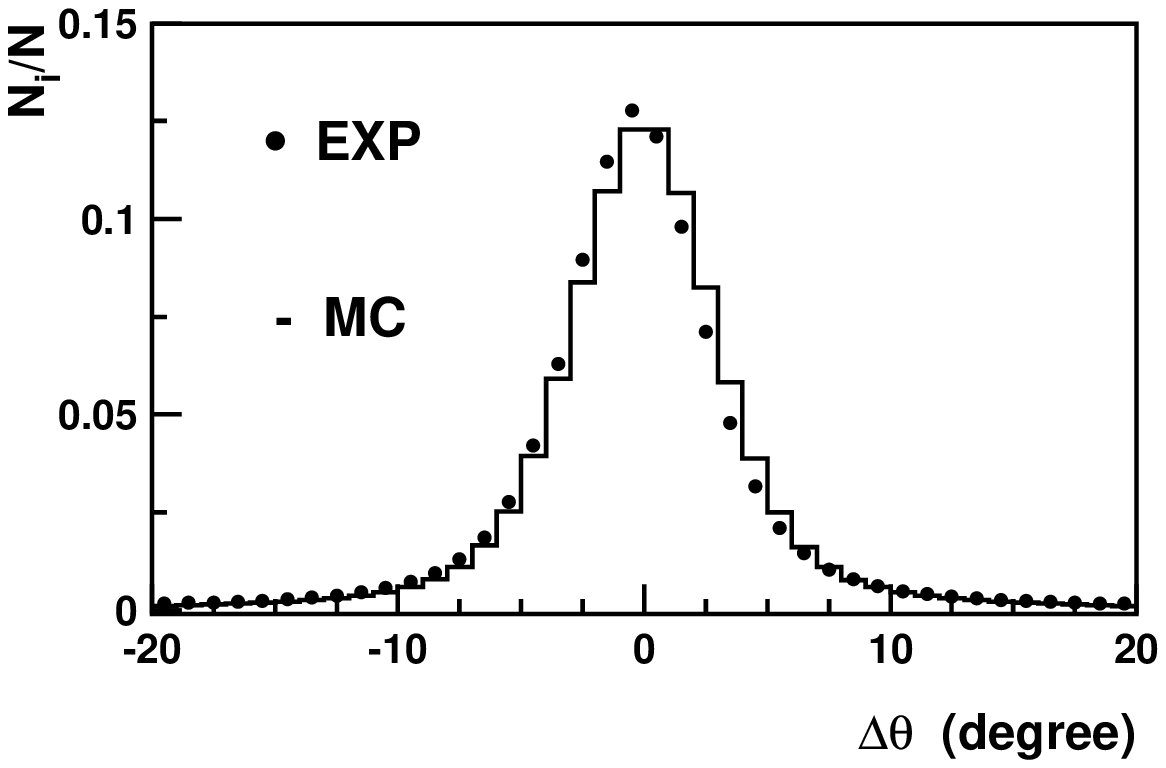,width=15.0cm}
\caption{The $\Delta\theta$ distribution of the $e^+e^-\to\pi^+\pi^-$
         events. Dots -- experiment, histogram -- simulation.}
\label{dtetpp}
\end{center}
\end{figure}

 The $\Delta\phi$ and $\Delta\theta$ distributions of the $e^+e^-\to e^+e^-$ 
 and $e^+e^-\to\pi^+\pi^-$ events are shown in Fig.\ref{dphiee},\ref{dphipp},
 \ref{dtetee} and \ref{dtetpp}. Experiment and simulation agree rather well.
 As a measure of systematic uncertainty due to  $\Delta\theta$ cut the
 following value was used:
\begin{eqnarray}
 \delta_{\Delta\theta} =
 {\delta^{\pi\pi}_{\Delta\theta} \over \delta^{ee}_{\Delta\theta}}, 
\end{eqnarray}
 where
$$
 \delta^{x}_{\Delta\theta}=
 {n_x(|\Delta\theta|<10^\circ)\over N_x(|\Delta\theta|<20^\circ)} \mbox{~} / 
 \mbox{~}
 {m_x(|\Delta\theta|<10^\circ)\over M_x(|\Delta\theta|<20^\circ)}, \mbox{~~}
 x=\pi\pi(ee).
$$
 Here $n_x(|\Delta\theta|<10^\circ)$ and $m_x(|\Delta\theta|<10^\circ)$ are
 the numbers of experimental and simulated events, selected under the
 condition $|\Delta\theta|<10^\circ$, while $N_x(|\Delta\theta|<20^\circ)$ and 
 $M_x(|\Delta\theta|<20^\circ)$ are the numbers of experimental and simulated
 events with $|\Delta\theta|<20^\circ$. The $\delta_{\Delta\theta}$ does not
 depend on energy, its average value is equal to 0.999 and it has systematic
 spread of 0.4 \%. This systematic 
 spread was added to the error of the cross section measurement in each
 energy point. Systematic error due to the $\Delta\phi$ cut is significantly
 lower and was neglected.

\begin{figure}
\begin{center}
\epsfig{figure=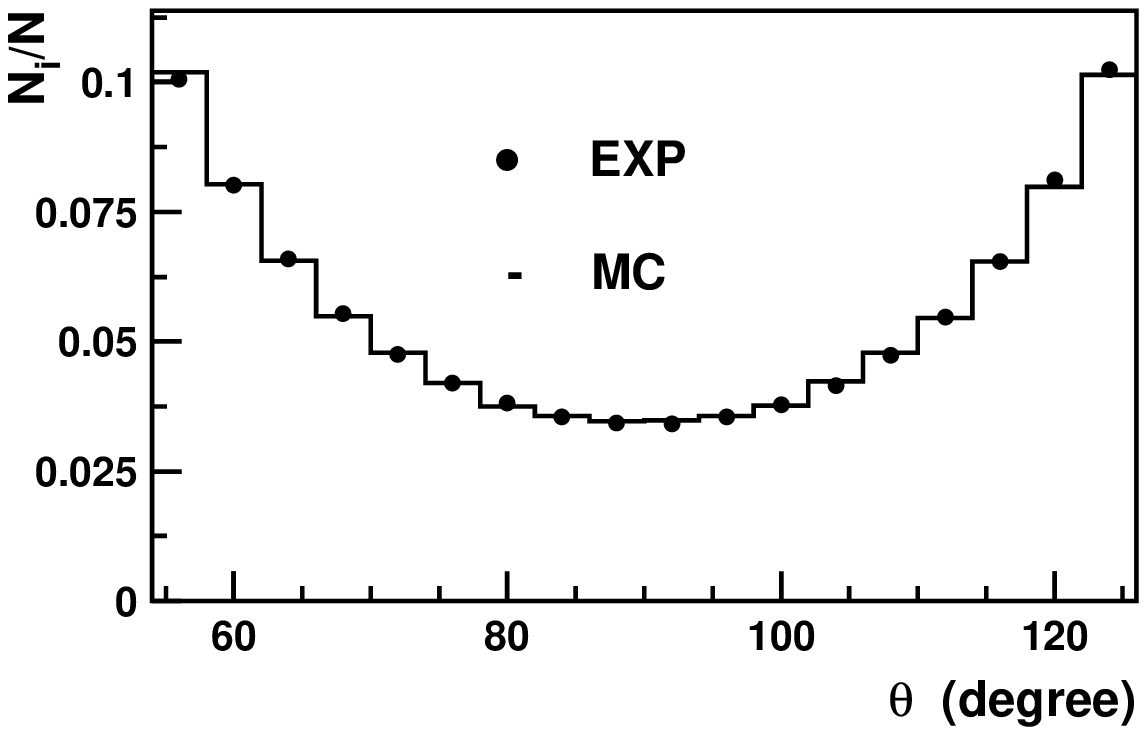,width=15.0cm}
\caption{The $\theta$ angle distribution of the $e^+e^-\to e^+e^-$
         events. Dots -- experiment, histogram -- simulation.}
\label{tetee}
\epsfig{figure=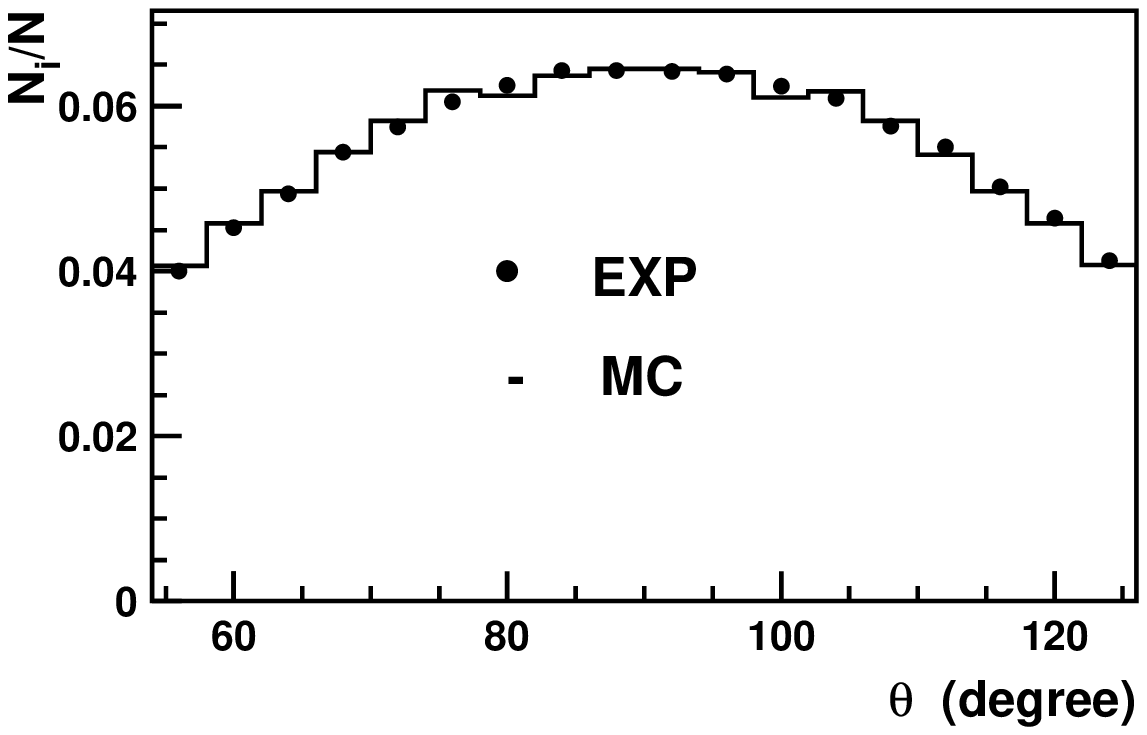,width=15.0cm}
\caption{The $\theta$ angle distribution of the $e^+e^-\to\pi^+\pi^-$
         events. Dots -- experiment, histogram -- simulation.}
\label{tetpp}
\end{center}
\end{figure}

\begin{figure}
\begin{center}
\epsfig{figure=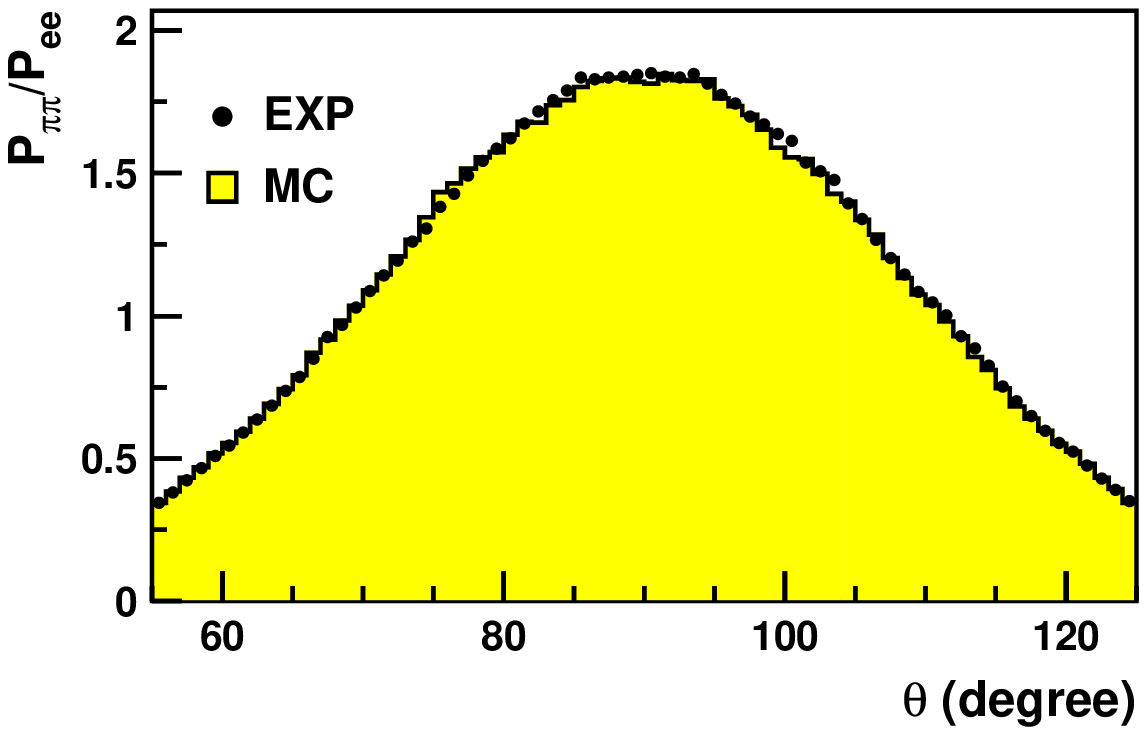,width=15.0cm}
\caption{The ratio of $\theta$ distributions of the $e^+e^-\to\pi^+\pi^-$
         and $e^+e^-\to e^+e^-$ 
	 processes. Dots -- experiment, histogram -- simulation.}
\label{epcpab}
\epsfig{figure=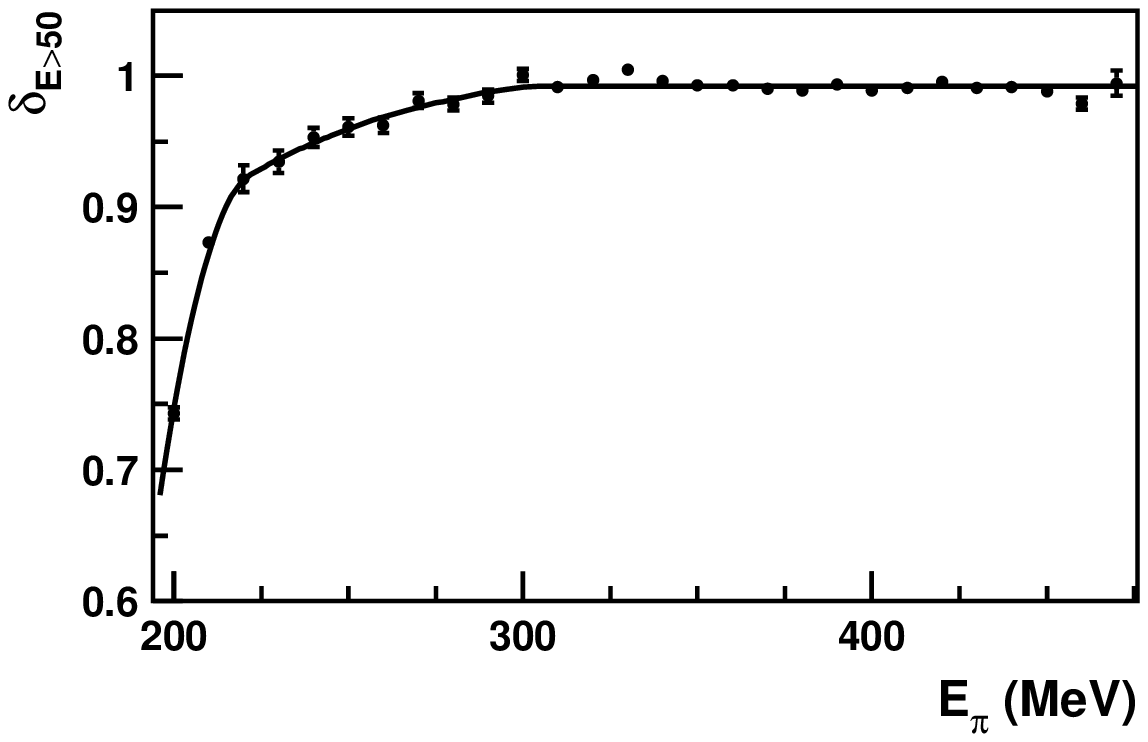,width=15.0cm}
\caption{The $\delta_{E>50}$ correction coefficient associated to the pions 
         energy deposition
         cut in dependence on the pion energy $E_\pi$.}
\label{non50}
\end{center}
\end{figure}

 The polar angle distributions for the $e^+e^-\to e^+e^-$ and
 $e^+e^-\to\pi^+\pi^-$ processes are shown in Fig.\ref{tetee} and
 \ref{tetpp}. The ratio of these $\theta$ distributions
 is shown in Fig.\ref{epcpab}. 
 The experimental and simulated distributions are in agreement. In order to
 estimate the systematic inaccuracy due to the $\theta$ angle selection cut
 the following ratio was used:
\begin{eqnarray}
 \delta_\theta={\delta(\theta_x)\over\delta(55^\circ)},
\end{eqnarray}
 where
$$
 \delta(\theta_x)=
 {N_{\pi\pi}(\theta_x<\theta<180^\circ-\theta_x) \over
  N_{ee}(\theta_x<\theta<180^\circ-\theta_x)} /
  {M_{\pi\pi}(\theta_x<\theta<180^\circ-\theta_x) \over
   M_{ee}(\theta_x<\theta<180^\circ-\theta_x)}, \mbox{~~} 
   50^\circ<\theta_x<90^\circ.
$$
 Here $N_{\pi\pi}(\theta_x<\theta<180^\circ-\theta_x)$,
     $N_{ee}(\theta_x<\theta<180^\circ-\theta_x)$,
     $M_{\pi\pi}(\theta_x<\theta<180^\circ-\theta_x)$,
     $M_{ee}(\theta_x<\theta<180^\circ-\theta_x)$ 
 are the experimental and simulated  $e^+e^-\to\pi^+\pi^-$ and  
 $e^+e^-\to e^+e^-$ event numbers 
 in the angular range $\theta_x<\theta<180^\circ-\theta_x$. 
 The maximal difference of $\delta_\theta$ from unity was found to be 0.8\%.
 This value was taken as a systematic error $\sigma_{\theta}=0.8$\% 
 associated with the angular selection cut.

 In the tracking system the particle track can be lost due to reconstruction
 inefficiency. The probabilities to find the track was determined by
 using experimental data themselves. It was found to be 
 $\varepsilon_{e}\simeq 0.996$ for electrons and  
 $\varepsilon_{\pi}\simeq 0.995$ for pions. In simulation these values 
 actually do not differ from unity, while in reality the track finding 
 probability for electrons is slightly greater then for pions. So the 
 detection efficiency was multiply by the correction coefficient:
\begin{eqnarray}
 \delta_{rec} =
  \Biggl[ {\varepsilon_{\pi} \over \varepsilon_{e}} \Biggr]^2 = 0,997
\end{eqnarray}

 Pions can be lost due to the nuclear interaction in the detector 
 material before the tracking system, for example, via the reaction 
 $\pi^\pm N\to\pi^\pm N$ with the final pion scattered at the large angle 
 or via charge exchange reaction $\pi^\pm N\to\pi^0 N$.
 As a measure of systematic inaccuracy associated to this effect the 
 difference from unity of the following quantity was used:
\begin{eqnarray}
 \delta_{nucl} = 
 \Biggl[ \biggl(1 - {n \over 3N}\biggr) / 
         \biggl(1 - {m \over 3M}\biggr) \Biggr]^2,
\end{eqnarray}
 where $N$ and $M$ is the pions numbers in experiment and simulation; $n$ and
 $m$ is the pions numbers in experiment and simulation which had a track
 in the drift chamber nearest to the beam-pipe, but the corresponding track 
 in the second drift chamber and associated cluster in the calorimeter were 
 not found. The  particle loss probability
 was divided by 3 -- the ratio of amounts of the matter 
 between the drift chambers and before the tracking system. The deviation of
 $\delta_{nucl}$ from 1 was taken as a systematic error $\sigma_{nucl}=0.2$
 \%.
 
 Uncertainties in simulation of pions nuclear interactions imply that the cut
 on the particles energy deposition leads to an inaccuracy in detection 
 efficiency of the $e^+e^-\to\pi^+\pi^-$ process. In order to take
 into account this inaccuracy, the detection efficiency was multiplied by the 
 correction coefficients. The correction coefficients was obtained by using
 events of the $e^+e^-\to\pi^+\pi^-\pi^0$ reaction
 \cite{phi98,pi3omeg,dplphi98}. Pions energies in the
 $e^+e^-\to\pi^+\pi^-\pi^0$ events were determined via the kinematic fit.
 The pion energies were divided into the 10 MeV wide bins . For each
 bin the correction coefficient (Fig.\ref{non50}) was obtained:
\begin{eqnarray}
 \delta_{E>50} = \biggl[{ {n_i/N_i} \over {m_i/M_i} }\biggr]^2,
\end{eqnarray}
 where $i$ is the bin number, $N_i$ and $M_i$ are the pions numbers in 
 experiment and simulation selected in the $i$th bin by the kinematic fit 
 without any cut on the energy deposition in the
 calorimeter ; $n_i$ and $m_i$ are the pions numbers in 
 experiment and simulation under the condition that the pion energy 
 deposition is greater than 50 MeV. To estimate systematic errors in 
 determination of these correction coefficients, the ratio of the probability
 that both pions in simulated $e^+e^-\to\pi^+\pi^-$  events have energy 
 deposition more than 50 MeV to the quantity $(m_i/M_i)^2$ was consider. 
 This ratio is 0.994 at
 $\sqrt{s}>420$ MeV and about 0.97 at $\sqrt{s}<420$ MeV. The difference
 of this ratio from unity was taken as a systematic error $\sigma_{E>50}$ of
 the $\delta_{E>50}$  correction coefficient determination: 
 $\sigma_{E>50}=0.6$ \% at
 $\sqrt{s}>420$ MeV and $\sigma_{E>50}=3$ \% at $\sqrt{s}<420$ MeV.
 
 In the energy region $\sqrt{s} = 840$ -- $970$ MeV the probability to hit
 the muon/veto system for muons and pions varies from 1\% upto 93\%, and
 from 0.5\% to 3\% respectively. The usage of the muon system $veto$ for 
 events selection ($veto=0$) leads to inaccuracy in the measured cross section 
 determination due to the uncertainty in the simulation of the muons and pions 
 traversing through
 the detector at $\sqrt{s}>840$ MeV. In order to obtain the necessary
 corrections, the events close to the
 median plane $\phi<10^\circ$, $170^\circ\phi<190^\circ$, $\phi>350^\circ$,
 where the cosmic background is minimal, were used.
 The $e^+e^-\to\pi^+\pi^-$ cross section was measured with ($veto=0$) and 
 without ($veto\ge 0$) using the muon system, and the following
 correction coefficient was obtained for each energy point:
\begin{eqnarray}
 \delta_{veto} ={ \sigma(e^+e^-\to\pi^+\pi^-;veto\ge 0) \over
                 \sigma(e^+e^-\to\pi^+\pi^-;veto  = 0)}
\end{eqnarray}
 It was found that $\delta_{veto}=0.95$ at $\sqrt{s}=970$ MeV and quickly
 rises up to 1 for lower energies.
 
\begin{figure}
\begin{center}
\epsfig{figure=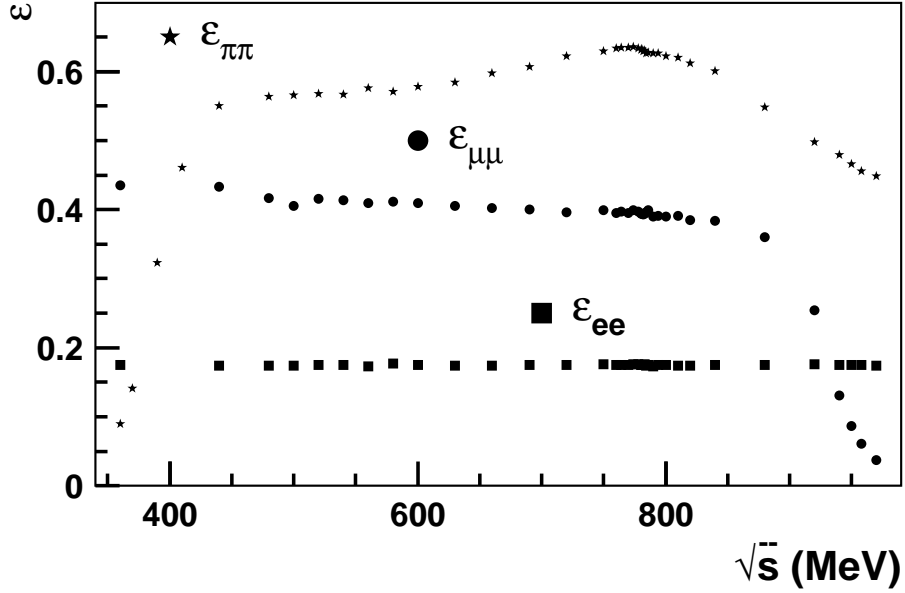,width=15.0cm}
\caption{The detection efficiencies $\varepsilon_{\pi\pi}$,
         $\varepsilon_{ee}$, $\varepsilon_{\mu\mu}$, of the
	 $e^+e^-\to\pi^+\pi^-, \mu^+\mu^-$ and
         $e^+e^-$ processes.}
\label{eff_peak}
\end{center}
\end{figure}

 The detection efficiencies of the processes $e^+e^-\to\pi^+\pi^-$,
 $\mu^+\mu^-$ and $e^+e^-$ after all applied corrections are shown in 
 Fig.\ref{eff_peak}. The detection efficiency of the $e^+e^-\to e^+e^-$
 reaction does not depend on energy, while  for $e^+e^-\to\mu^+\mu^-$ and 
 $\pi^+\pi^-$ processes it does. The decrease  of the 
 $e^+e^-\to\mu^+\mu^-$ process detection efficiency at 
 $\sqrt{s}>800$ MeV is caused by the fact that
 the probability for muons to hit the muon system
 rises with energy. The detection efficiency of the
 $e^+e^-\to\pi^+\pi^-$ process at $\sqrt{s}>500$ MeV is determined
 mainly by the cuts on the pions angles. Below 500 MeV the detection 
 efficiency 
 decreases due to the cut on the pions energy deposition in the calorimeter.
 The statistical error $\le 1\%$ of the detection efficiency determination
 was added to the cross section measurement error in each energy point.
 The total systematic error of the detection efficiency determination
 $\sigma_{eff}=\sigma_{E>50}\oplus\sigma_{nucl}\oplus\sigma_{\theta}$
 is $\sigma_{eff}=1$ \% at $\sqrt{s}\ge 420$ MeV and
 $\sigma_{eff}=3.1$ \% at $\sqrt{s}<420$ MeV.
 
\subsection{Measurement of the $e^+e^-\to\pi^+\pi^-$ cross section.}

\begin{figure}
\begin{center}
\epsfig{figure=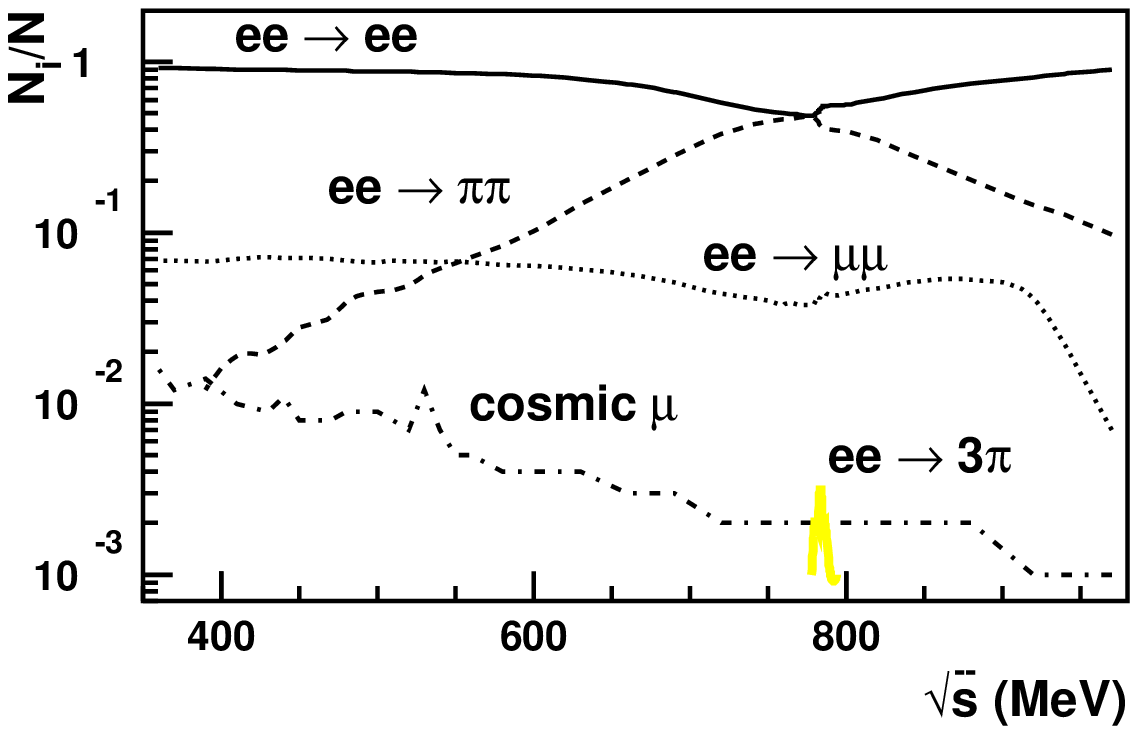,width=15.0cm}
\caption{The percentage of the $e^+e^-\to e^+e^-,
         \pi^+\pi^-,\mu^+\mu^-, \pi^+\pi^-\pi^0$ and
         cosmic background events in dependence on energy $\sqrt{s}$}
\label{doli}
\epsfig{figure=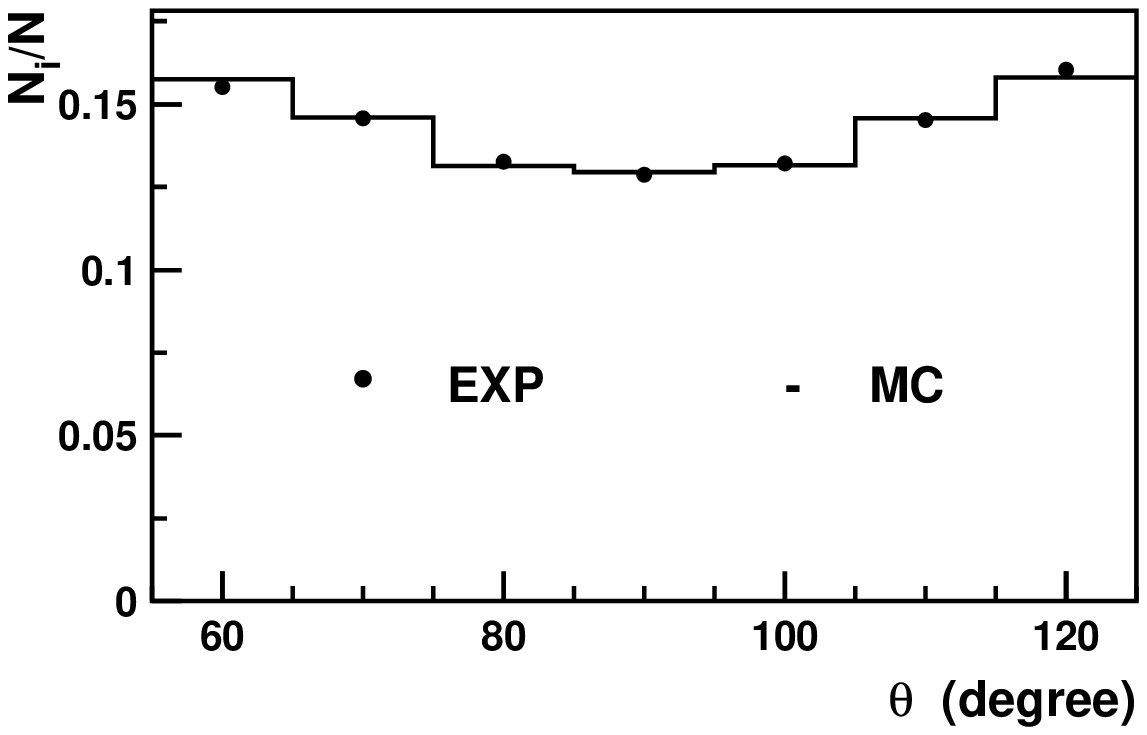,width=15.0cm}
\caption{The $\theta$ angle distributions of all collinear events at
         $\sqrt{s}$ from 880 MeV to 630 MeV.
	 Dots -- experiment, histogram -- simulation.}
\label{tetbce2}
\end{center}
\end{figure}

 The number of selected events in the regions $R_{e/\pi}<0.5$ and
 $R_{e/\pi}>0.5$ are:
\begin{eqnarray}
 N=N_{\pi\pi}+N_{ee}+N_{\mu\mu}+N_{\mu}+N_{3\pi},
\end{eqnarray}
\begin{eqnarray}
 M=M_{\pi\pi}+M_{ee}+M_{\mu\mu}+M_{\mu}+M_{3\pi}.
\end{eqnarray}
 Here $N$ and $M$ are the events numbers in the regions $R_{e/\pi}<0.5$ and
 $R_{e/\pi}>0.5$ respectively. $N_{\mu}$, $M_{\mu}$ and $N_{3\pi}$, 
 $M_{3\pi}$ are the number of background events due to cosmic muons and the 
 $e^+e^-\to\pi^+\pi^-\pi^0$ process, calculated as was described above. The
 $e^+e^-\to\mu^+\mu^-$ process  events number can be written as:
\begin{eqnarray}
 N_{\mu\mu}=\sigma_{\mu\mu}\times\varepsilon_{\mu\mu}\times
 (1-\epsilon_{\mu\mu})\times IL,
\end{eqnarray}
\begin{eqnarray}
 M_{\mu\mu}=\sigma_{\mu\mu}\times\varepsilon_{\mu\mu}\times
 \epsilon_{\mu\mu}\times IL,
\end{eqnarray}
 where $\sigma_{\mu\mu}$ is the $e^+e^-\to\mu^+\mu^-$ process
 cross section obtained according to Ref.\cite{arbuzqed}, 
 $\varepsilon_{\mu\mu}$ is the process detection efficiency, 
 $\epsilon_{\mu\mu}$ is the probability for the $e^+e^-\to\mu^+\mu^-$ process
 events to have $R_{e/\pi}>0.5$.
 $IL$ is the integrated luminosity:
\begin{eqnarray}
 IL = {M_{ee} \over {\sigma_{ee }\times\varepsilon_{ee }\times
 \epsilon_{ee }}},
\end{eqnarray}
 where $\varepsilon_{ee }$ and $\epsilon_{ee }$ are the detection
 efficiency and the probability to have $R_{e/\pi}>0.5$ for the process 
 $e^+e^-\to e^+e^-$, $\sigma_{ee }$ is the process cross section 
 with the $30^\circ<\theta<150^\circ$ angular cut for the electron and positron
 in the final state. The cross section 
 $\sigma_{ee }$ was calculated by using BHWIDE 1.04 \cite{bhwide} code with 
 accuracy  0.5 \%. The $e^+e^-\to\pi^+\pi^-$ process events number 
 with $R_{e/\pi}>0.5$ and the $e^+e^-\to e^+e^-$ process events number with 
 $R_{e/\pi}<0.5$ can be written in the following way:
$$
 N_{ee}={{1-\epsilon_{ee }} \over \epsilon_{ee }} \times M_{ee}
       =\lambda_{ee} \times M_{ee}, \mbox{~~~} 
 M_{\pi\pi}= {{1-\epsilon_{ee }} \over \epsilon_{ee }}
             \times N_{\pi\pi}
           = \lambda_{\pi\pi}\times N_{\pi\pi}.
$$
 The  $e^+e^-\to e^+e^-$ process events number with $R_{e/\pi}>0.5$
 and the $e^+e^-\to\pi^+\pi^-$ process events number with
 $R_{e/\pi}<0.5$ are equal to:
\begin{eqnarray}
 M_{ee}={{M-M_\mu-\lambda_{\pi\pi} \times (N-N_{\mu})} \over 
         {\kappa - \Delta \times \lambda_{\pi\pi}}},
\end{eqnarray}
\begin{eqnarray}
 N_{\pi\pi}=N-N_{\mu}-M_{ee}\times \Delta.
\end{eqnarray}
 Here
$$
 \Delta = \lambda_{ee} + 
 {{\sigma_{\mu\mu}\times\varepsilon_{\mu\mu}\times
 (1-\epsilon_{\mu\mu})+N_{3\pi}/IL} \over
 {\sigma_{ee}\times\varepsilon_{ee}\times\epsilon_{ee}}},
$$
$$
 \kappa = 1 +
 {{\sigma_{\mu\mu}\times\varepsilon_{\mu\mu}\times
 \epsilon_{\mu\mu}+M_{3\pi}/IL} \over
 {\sigma_{ee}\times\varepsilon_{ee}\times\epsilon_{ee}}}.
$$
 The percentage of each process in the selected events in dependence on
 energy $\sqrt{s}$ is shown in Fig.\ref{doli}. The experimental angular
 distributions agree with the sum of distributions for each process weighted
 according to its contribution (Fig.\ref{tetbce2}).
  
 The $e^+e^-\to\pi^+\pi^-$ process cross section is calculated
 from the following formula:
\begin{eqnarray}
 \sigma_{\pi\pi} = {N_{\pi\pi} \over 
 {IL \times\varepsilon_{\pi\pi}\times(1-\epsilon_{\pi\pi})}} =
 {{\sigma_{ee}\times\varepsilon_{ee}\times\epsilon_{ee}}
 \over
 {\varepsilon_{\pi\pi}\times(1-\epsilon_{\pi\pi})}} \times
 \Biggl[{ {\kappa-\Delta\times\lambda_{\pi\pi}} \over 
 { { {M-M_{\mu}} \over {N-N_{\mu}} }-\lambda_{\pi\pi}} }-\Delta \Biggr].
\end{eqnarray}

\begin{figure}
\begin{center}
\epsfig{figure=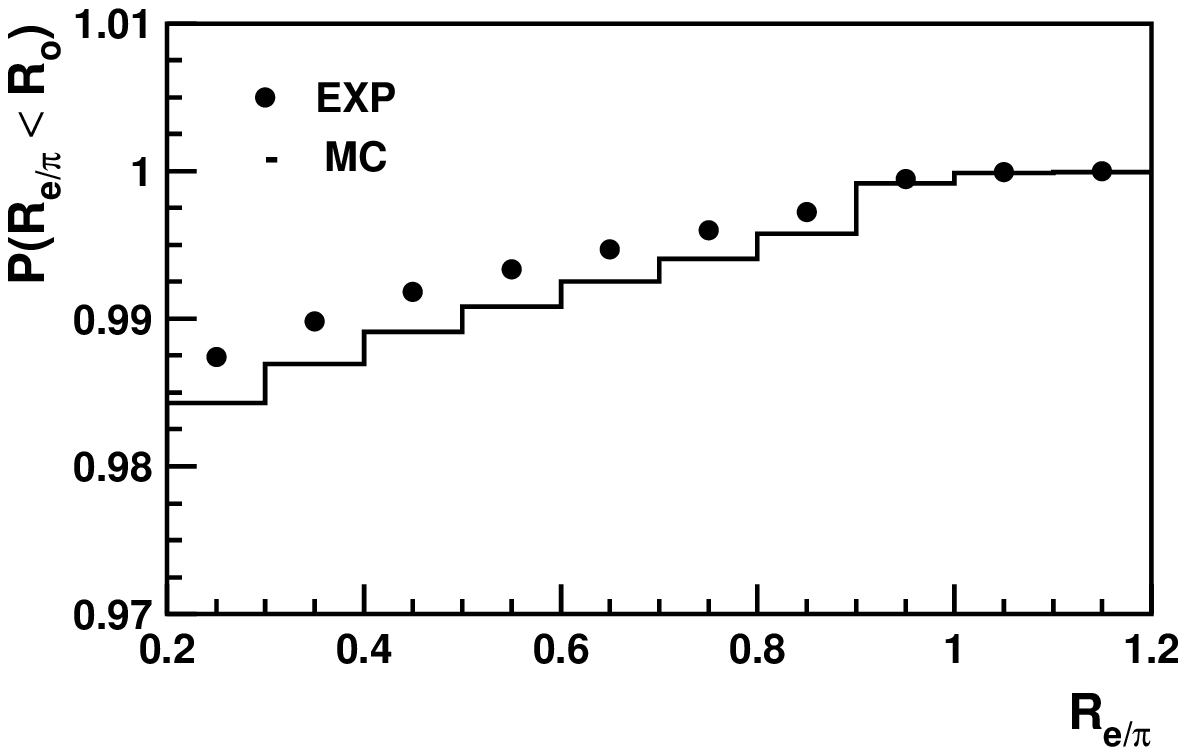,width=15.0cm}
\caption{The probability of the pseudo $\pi\pi$ events to have  $R_{e/\pi}$
         value less than some $R_0$.
	 Dots -- experiment, histogram -- simulation.}
\label{mlp_nce_nn}
\epsfig{figure=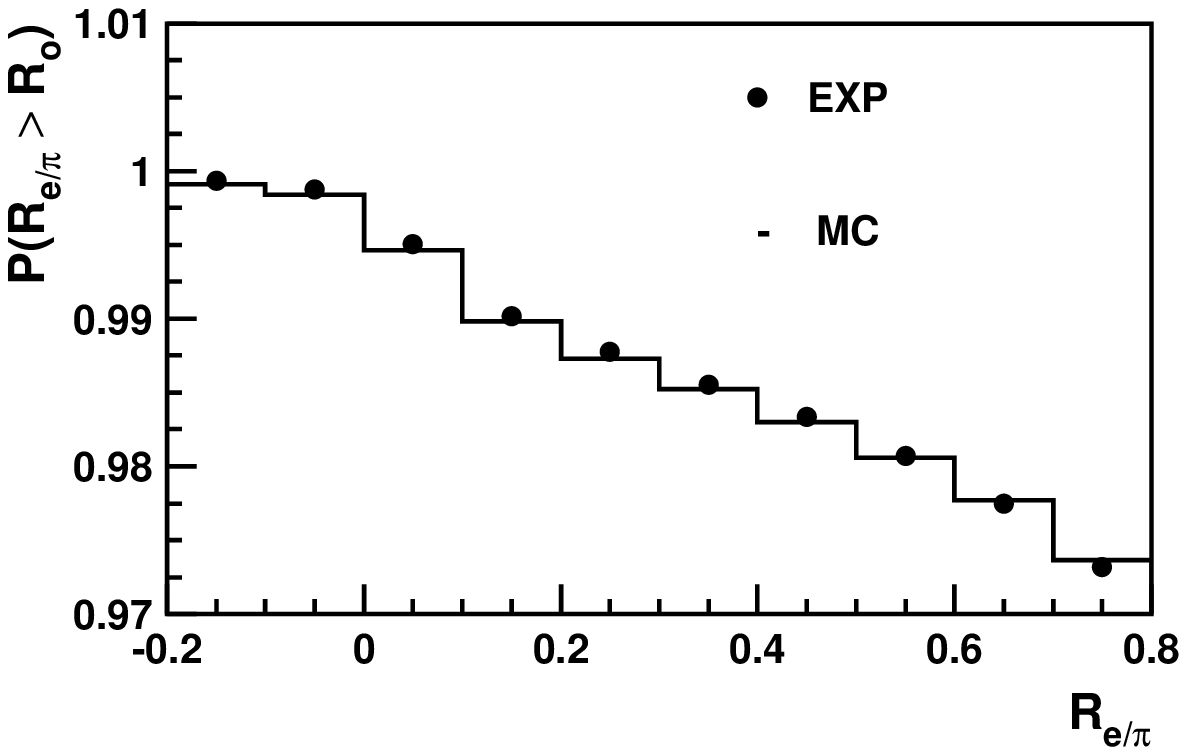,width=15.0cm}
\caption{The probability of the pseudo $ee$ events to have $R_{e/\pi}$
         value greater than some $R_0$.
	 Dots -- experiment, histogram -- simulation.}
\label{mlp_nce_ee}
\end{center}
\end{figure}

 In order to estimate the systematic uncertainty due to $e-\pi$ 
 discrimination, the pseudo $\pi\pi$ and pseudo $ee$ events in the experiment 
 and simulation were formed. The pseudo $\pi\pi$ events were constructed by
 using pions from the $e^+e^-\to\pi^+\pi^-\pi^0$ reaction. 
 In order to construct
 the pseudo $\pi\pi$ event with the pions having energy $E_0$, two charged 
 pions with energies $E_\pi$ such that $|E_0-E_\pi|<10$ MeV were used from
 two separate $e^+e^-\to\pi^+\pi^-\pi^0$ events. 
 Of course, such pseudo 
 $\pi\pi$ events are in general not collinear but this is irrelevant for our
 purposes here.
 The pseudo $ee$ event was constructed analogously from the particles of
 two separate collinear events such that their partners in these events have
 energy depositions in the calorimeter layers
 typical for electrons.  Fig.\ref{mlp_nce_nn} and \ref{mlp_nce_ee} show
 probabilities for the discrimination parameter to have values 
 less than some magnitude in experiment and simulation for such pseudo events. 
 Using these distributions, the corrections to the probabilities for 
 the separation parameter $R_{e/\pi}$ to be greater or less than 0.5 was 
 obtained. 
 The difference between cross sections measured with and without
 these corrections was taken as a systematic error and its
 value does not exceed 0.5 \% for different energy points.

\begin{table}
\begin{center}
\caption{The results of the $e^+e^-\to\pi^+\pi^-$ cross section measurements.
         $\sigma_{\pi\pi}$ is the 
	 $e^+e^-\to\pi^+\pi^-$ cross section taking into account the 
	 radiative corrections due to the initial and final state radiation,
         $\delta_{rad}$ is the radiative correction due to the initial and 
         final state radiation, $\sigma_0$ and
	 $|F_\pi|^2$ are the cross section and the  form factor of the 
	 $e^+e^-\to\pi^+\pi^-$ process after the radiative corrections were
	 undressed, $\sigma^{pol}_{\pi\pi}$ is the
	 $e^+e^-\to\pi^+\pi^-$ undressed cross section without vacuum 
         polarization but with the final state radiation. 
         Only uncorrelated errors are shown.
	 The correlated systematic error $\sigma_{sys}$ is 1.3 \% for 
	 $\sqrt{s}\ge 420$ MeV and 3.2 \% for $\sqrt{s}< 420$ MeV.}
\label{tab1}
\begin{tabular}[t]{cccccc}
$\sqrt[]{s}$ (MeV)&$\sigma_{\pi\pi}$(nb) &
$\delta_{rad}$&$\sigma_0$ (nb)&$|F_\pi|^2$&
$\sigma^{pol}_{\pi\pi}$(nb) \\ \hline
970.& 118.12$\pm$ 2.76&1.491&  79.20$\pm$ 1.85&  3.91$\pm$ 0.09&   77.53$\pm$
  1.81 \\
958.& 137.16$\pm$ 2.94&1.454&  94.34$\pm$ 2.02&  4.56$\pm$ 0.10&   92.16$\pm$
  1.97 \\
950.& 150.02$\pm$ 2.85&1.430& 104.88$\pm$ 1.99&  4.99$\pm$ 0.09&  102.35$\pm$
  1.94 \\
940.& 166.55$\pm$ 2.27&1.400& 119.00$\pm$ 1.62&  5.56$\pm$ 0.08&  116.01$\pm$
  1.58 \\
920.& 204.99$\pm$ 7.14&1.340& 152.96$\pm$ 5.33&  6.89$\pm$ 0.24&  148.60$\pm$
  5.18 \\
880.& 310.82$\pm$ 3.52&1.220& 254.67$\pm$ 2.88& 10.65$\pm$ 0.12&  245.94$\pm$
  2.78 \\
840.& 513.80$\pm$ 4.76&1.106& 464.48$\pm$ 4.30& 17.99$\pm$ 0.17&  446.64$\pm$
  4.13 \\
820.& 676.03$\pm$ 5.99&1.055& 640.60$\pm$ 5.68& 23.86$\pm$ 0.21&  614.57$\pm$
  5.45 \\
810.& 760.19$\pm$ 6.58&1.032& 736.34$\pm$ 6.37& 26.90$\pm$ 0.23&  704.79$\pm$
  6.10 \\
800.& 856.66$\pm$ 7.32&1.013& 845.61$\pm$ 7.23& 30.28$\pm$ 0.26&  807.33$\pm$
  6.90 \\
794.& 890.86$\pm$ 7.43&1.009& 883.09$\pm$ 7.37& 31.25$\pm$ 0.26&  838.38$\pm$
  7.00 \\
790.& 892.35$\pm$17.70&1.015& 879.09$\pm$17.44& 30.86$\pm$ 0.61&  829.16$\pm$
 16.45 \\
786.& 926.47$\pm$ 7.84&1.031& 898.19$\pm$ 7.60& 31.28$\pm$ 0.26&  842.92$\pm$
  7.13 \\
785.& 941.34$\pm$ 9.33&1.032& 911.99$\pm$ 9.04& 31.70$\pm$ 0.31&  858.12$\pm$
  8.51 \\
784.& 989.76$\pm$20.12&1.025& 966.05$\pm$19.64& 33.51$\pm$ 0.68&  915.22$\pm$
 18.61 \\
783.&1060.12$\pm$11.38&1.010&1050.08$\pm$11.27& 36.35$\pm$ 0.39& 1005.99$\pm$
 10.80 \\
782.&1123.55$\pm$26.83&0.989&1136.34$\pm$27.14& 39.26$\pm$ 0.94& 1102.62$\pm$
 26.33 \\
781.&1158.03$\pm$10.80&0.971&1192.83$\pm$11.12& 41.13$\pm$ 0.38& 1169.48$\pm$
 10.90 \\
780.&1211.67$\pm$ 9.98&0.957&1266.56$\pm$10.43& 43.59$\pm$ 0.36& 1252.62$\pm$
 10.32 \\
778.&1273.38$\pm$ 9.47&0.944&1349.27$\pm$10.03& 46.25$\pm$ 0.34& 1343.80$\pm$
  9.99 \\
774.&1282.06$\pm$ 9.49&0.938&1366.85$\pm$10.12& 46.48$\pm$ 0.34& 1361.99$\pm$
 10.08 \\
770.&1249.25$\pm$ 9.26&0.935&1336.51$\pm$ 9.91& 45.08$\pm$ 0.33& 1330.42$\pm$
  9.86 \\
764.&1247.24$\pm$ 9.35&0.932&1338.62$\pm$10.04& 44.61$\pm$ 0.33& 1331.35$\pm$
  9.99 \\
760.&1244.74$\pm$ 9.58&0.927&1342.60$\pm$10.33& 44.39$\pm$ 0.34& 1335.30$\pm$
 10.27 \\
750.&1219.07$\pm$21.50&0.920&1325.56$\pm$23.38& 42.95$\pm$ 0.76& 1321.82$\pm$
 23.31 \\
720.& 989.95$\pm$ 6.62&0.910&1087.59$\pm$ 7.27& 33.15$\pm$ 0.22& 1091.88$\pm$
  7.30 \\
690.& 717.99$\pm$ 7.78&0.915& 784.79$\pm$ 8.50& 22.50$\pm$ 0.24&  789.95$\pm$
  8.56 \\
660.& 515.95$\pm$ 5.87&0.923& 558.83$\pm$ 6.36& 15.07$\pm$ 0.17&  561.19$\pm$
  6.39 \\
630.& 382.69$\pm$ 8.35&0.933& 410.32$\pm$ 8.95& 10.41$\pm$ 0.23&  411.22$\pm$
  8.97 \\
600.& 287.18$\pm$10.56&0.940& 305.50$\pm$11.23&  7.30$\pm$ 0.27&  305.61$\pm$
 11.23 \\
580.& 255.24$\pm$14.39&0.945& 270.24$\pm$15.24&  6.22$\pm$ 0.35&  269.85$\pm$
 15.22 \\
560.& 226.60$\pm$12.41&0.948& 239.01$\pm$13.09&  5.30$\pm$ 0.29&  238.63$\pm$
 13.07 \\
550.& 217.52$\pm$17.51&0.950& 228.99$\pm$18.43&  4.99$\pm$ 0.40&  228.29$\pm$
 18.37 \\
540.& 212.67$\pm$13.55&0.952& 223.47$\pm$14.24&  4.78$\pm$ 0.30&  222.82$\pm$
 14.20 \\
530.& 200.04$\pm$22.75&0.953& 210.00$\pm$23.88&  4.42$\pm$ 0.50&  209.43$\pm$
 23.82 \\
520.& 178.13$\pm$10.25&0.954& 186.73$\pm$10.75&  3.87$\pm$ 0.22&  186.26$\pm$
 10.72 \\
510.& 174.28$\pm$16.65&0.954& 182.60$\pm$17.45&  3.73$\pm$ 0.36&  181.82$\pm$
 17.38 \\
500.& 175.22$\pm$10.78&0.955& 183.52$\pm$11.29&  3.70$\pm$ 0.23&  182.77$\pm$
 11.24 \\
480.& 165.18$\pm$ 9.58&0.955& 172.90$\pm$10.03&  3.41$\pm$ 0.20&  172.29$\pm$
  9.99 \\
470.& 143.94$\pm$13.21&0.955& 150.71$\pm$13.83&  2.94$\pm$ 0.27&  150.22$\pm$
 13.78 \\
450.& 141.32$\pm$14.21&0.954& 148.10$\pm$14.89&  2.86$\pm$ 0.29&  147.42$\pm$
 14.82 \\
440.& 116.15$\pm$15.58&0.953& 121.86$\pm$16.35&  2.35$\pm$ 0.32&  121.34$\pm$
 16.28 \\
430.& 111.27$\pm$12.60&0.952& 116.86$\pm$13.23&  2.26$\pm$ 0.26&  116.41$\pm$
 13.18 \\
410.& 127.38$\pm$19.11&0.949& 134.23$\pm$20.14&  2.64$\pm$ 0.40&  133.84$\pm$
 20.08 \\
390.& 121.81$\pm$22.48&0.944& 128.98$\pm$23.80&  2.65$\pm$ 0.49&  128.76$\pm$
 23.76 \\
\hline
\end{tabular}
\end{center}
\end{table}

\begin{table}[ccch]
\begin{center}
\caption{Various contributions to the systematic error of the 
         $e^+e^-\to\pi^+\pi^-$
         cross section determination. $\sigma_{sys}$ is the total systematic
         error, $\sigma_{eff}=\sigma_{E>50}\oplus\sigma_{nucl}\oplus
         \sigma_{\theta}$
	 is the systematic inaccuracy of the detection efficiency 
         determination.}
\label{tabcuc}
\begin{tabular}[t]{lcc}
 Error&Contribution at $\sqrt[]{s}\ge 420$ MeV&Contribution at
 $\sqrt[]{s}<420$ MeV \\ \hline
 $\sigma_{E>50}$&0.6 \%&3.0 \% \\
 $\sigma_{nucl}$&0.2 \%&0.2 \% \\
 $\sigma_{\theta}$&0.8 \%&0.8 \% \\ \hline
 $\sigma_{eff}$&1.0 \%&3.1 \% \\
 $\sigma_{sep}$&0.5 \%&0.5 \% \\
 $\sigma_{IL}$&0.5 \%&0.5 \% \\
 $\sigma_{rad}$&0.2 \%&0.2 \% \\
\hline
 $\sigma_{sys}$&1.3 \%&3.2 \% \\
\hline
\end{tabular}
\end{center}
\end{table}
 
 The obtained cross sections together with the radiative corrections 
 $\delta_{rad}$, including the
 initial and final state radiation, are presented in Table~\ref{tab1}.
 The $\delta_{rad}$ radiative correction was calculated according to 
 Ref.\cite{arbuzhad}.
 The accuracy of its determination is 0.2 \%. Having at hand the radiative 
 corrections the Born cross section for
 the $e^+e^-\to\pi^+\pi^-$ process can be extracted as
 follows
\begin{eqnarray}
 \sigma_0(s) = {\sigma_{\pi\pi}(s) \over \delta_{rad}(s)}
\end{eqnarray}
 The value of $\delta_{rad}(s)$ depends on the cross section at lower 
 energies, so it was calculated iteratively. The iteration stops then its 
 value changes by not more than  0.1 \% in consecutive iterations.
 The form factor values
 $$|F_\pi(s)|^2={{3s}\over{\pi\alpha^2\beta^3}}\sigma_{\pi\pi}(s),
 \mbox{~~} \beta=\sqrt{1-4m_{\pi}^2/s}$$
 are also listed in Table~\ref{tab1}. To evaluate the value of
 $R(s)=\sigma(e^+e^-\to\mbox{hadrons})/\sigma(e^+e^-\to\mu^+\mu^-)$,
 which is used in dispersion integrals calculation, the bare cross section
 $e^+e^-\to\pi^+\pi^-$ is used (the cross section without vacuum polarization
 contribution but taking into account the final state radiation):
\begin{eqnarray}
 \sigma^{pol}_{\pi\pi}(s)=\sigma_0(s)\times  |1-\Pi(s)|^2 \times
 \biggl(1+{\alpha\over\pi}a(s)\biggr),
\end{eqnarray}
 where $\Pi(s)$ is the polarization operator calculated according to the 
 Ref.\cite{arbuzqed} from the known $e^+e^-\to\mbox{hadrons}$ cross section 
 \cite{fedor}. 
 The last factor takes into account the final state radiation, and $a(s)$
 has the form \cite{shw}
$$
 a(s)={1+\beta^2 \over \beta}\biggl[ 
 4Li_2\biggl({1-\beta \over 1+\beta}\biggr) +
 2Li_2\biggl(-{1-\beta \over 1+\beta}\biggr)-
 3\ln{2\over 1+\beta}\ln{1+\beta \over 1-\beta} -
 2\ln{\beta}\ln{1+\beta \over 1-\beta} \biggr] - 
$$
$$
 - 3\ln{4\over 1-\beta^2} - 4\ln{\beta} +
 {1\over\beta^3}\biggl[ {5\over 4}(1+\beta^2)^2-2 \biggr] \times
 \ln{1+\beta \over 1-\beta} + {3\over 2}{1+\beta^2\over\beta^2}.
$$
 Here
$$
 Li_2(x) = -\int\limits^x_0 dt\ln(1-t)/t.
$$
 The values of $\sigma^{pol}_{\pi\pi}(s)$ are listed in Table~\ref{tab1}.
 
 The total systematic error of the cross section determination is:
 $$
 \sigma_{sys}=\sigma_{eff}\oplus\sigma_{sep}\oplus\sigma_{IL}
 \oplus\sigma_{rad}.
 $$
 Here $\sigma_{eff}$ is the systematic error of the detection efficiency
 determination,
 $\sigma_{sep}$ is the systematic error associated with the $e-\pi$ 
 separation,
 $\sigma_{IL}$ is the systematic error of the integrated luminosity
 determination, and  $\sigma_{rad}$ is the uncertainty of the radiative 
 correction calculation.
 The magnitudes of various contributions to the total systematic error are 
 shown in Table~\ref{tabcuc}. The total systematic error of the cross section
 determinations is $\sigma_{sys}=1.3$ \% at $\sqrt{s}\ge 420$ MeV and
 $\sigma_{sys}=3.2$ \% at $\sqrt{s}<420$ MeV.

\section{The $e^+e^-\to\pi^+\pi^-$ cross section analysis}
\subsection{Theoretical framework}
 In the framework of the vector meson dominance model, the cross section
 of the  $e^+e^-\to\pi^+\pi^-$ process is
\begin{eqnarray}
\label{cspp}
 \sigma_{\pi\pi}(s)={4\pi\alpha^2\over s^{3/2}} P_{\pi\pi}(s)|A_{\pi\pi}(s)|^2
\end{eqnarray}
 Here $P_{\pi\pi}(s)$ is the phase space factor:
$$
 P_{\pi\pi}(s)= q^3_{\pi}(s), \mbox{~~~} q_\pi(s)={1\over 2}\sqrt{s-4m^2_\pi}.
$$
 Amplitudes of the  $\gamma^\star \to \pi^+\pi^-$ transition have the form:
\begin{eqnarray}
\label{amp2p}
 |A_{\pi\pi}(s)|^2 = \Biggl| \sqrt{3\over 2} {1\over\alpha}
 \sum_{V=\rho,\omega,\rho^\prime,\rho^{\prime\prime}}
 { {\Gamma_V m_V^3 \mbox{~} \sqrt[]{m_V\sigma(V\to\pi^+\pi^-)}} \over 
   {D_V(s)} }
 { {e^{i\phi_{\rho V}} \over {\sqrt[]{q^3_{\pi}(m_V)}}} }\Biggr|^2,
\end{eqnarray}
 where
$$D_V(s)=m_V^2-s-i\mbox{~}\sqrt[]{s}\Gamma_V(s), \mbox{~~~}
  \Gamma_V(s)=\sum_{f}\Gamma(V\to f,s).$$
 Here $f$ denotes the final state of the $V$ vector meson decay, $m_V$ is
 the vector meson mass, $\Gamma_V=\Gamma_V(m_V)$.  The following forms of the
 energy dependence of the vector mesons total widths were used:
$$ \Gamma_\omega(s) = {m_\omega^2 \over s}{q^3_\pi(s) \over q^3_\pi(m_\omega)}
   \Gamma_\omega B(\omega\to\pi^+\pi^-) +
   {q^3_{\pi\gamma}(s) \over q^3_{\pi\gamma}(m_\omega)}
   \Gamma_\omega B(\omega\to\pi^0\gamma) +
   {W_{\rho\pi}(s) \over W_{\rho\pi}(m_\omega)}
   \Gamma_\omega B(\omega\to 3\pi),
$$
$$
 \Gamma_V(s)={m_V^2 \over s}{q^3_\pi(s) \over q^3_\pi(m_V)}\Gamma_V
 \mbox{~~~} (V=\rho,\rho^\prime,\rho^{\prime\prime})
$$
 Here $q_{\pi\gamma}=(s-m^2_\pi)/2\sqrt{s}$, $W_{\rho\pi}(s)$ is the phase
 space factor for the $\rho\pi\to\pi^+\pi^-\pi^0$ final state 
 \cite{phi98,pi3omeg,dplphi98}. In the energy dependence of the 
 $\rho,\rho^\prime,\rho^{\prime\prime}$ mesons widths only the 
 $V\to\pi^+\pi^-$ decays were taken into account.
 Such approach is justified in the energy region $\sqrt{s}<1000$ MeV. 
 Nowadays the $\rho^\prime,\rho^{\prime\prime}$ decays are rather poorly known
 and therefore the same approximation was used also for 
 the fitting  of the data above 1000 MeV.
 The $\omega$-meson mass  and width were taken from the SND measurements:
 $m_{\omega}=782.79$ MeV, $\Gamma_{\omega}=8.68$ MeV \cite{pi3omeg}.
 
 The relative decay probabilities  were calculated as follows
 $$B(V\to X)={\sigma(V\to X)\over\sigma(V)}, \mbox{~~}
 \sigma(V)=\sum_{X} \sigma(V\to X),  \mbox{~~}
 \sigma(V\to X) = {{12\pi B(V\to e^+e^-)B(V\to X) } \over {m_V^2}}.$$
 In the analysis presented here we have used $\sigma(\omega\to\pi^0\gamma)=
 155.8$ nb,
 $\sigma(\omega\to 3\pi)=1615$ nb obtained in the SND experiments 
 \cite{pi3omeg,pi0gam}.
 
 The parameter $\phi_{\rho V}$ is the relative interference phase between the 
 vector mesons
 $V$ and $\rho$, so $\phi_{\rho\rho}=0$.
 The phases $\phi_{\rho V}$ can deviate from $180^\circ$ or $0^\circ$,
 and their values can be energy dependent due to mixing between vector 
 mesons. The phases $\phi_{\rho\rho^\prime}$ and
 $\phi_{\rho\rho^{\prime\prime}}$ were fixed at $180^\circ$ and $0^\circ$,
 because these values are consistent with the existing experimental data 
 for the $e^+e^-\to\pi^+\pi^-$ reaction.
  
 Taking into account the  $\rho-\omega$ mixing, the $\omega\to\pi^+\pi^-$ and
 $\rho\to\pi^+\pi^-$ transition amplitudes can be written in the following 
 way \cite{thrhoom,akozi} 
\begin{eqnarray}
 \label{rhoom}
 A_{\omega\to\pi^+\pi^-} + A_{\rho\to\pi^+\pi^-} =
 { {g^{(0)}_{\gamma\rho}g^{(0)}_{\rho\pi\pi}} \over {D_{\rho}(s)} }
 \biggl[1-{g^{(0)}_{\gamma\omega}\over g^{(0)}_{\gamma\rho}}\varepsilon(s) 
 \biggr]+
 {{g^{(0)}_{\gamma\omega}g^{(0)}_{\rho\pi\pi}}\over{D_{\omega}(s)}}
 \biggl[\varepsilon(s)+{g^{(0)}_{\omega\pi\pi}\over g^{(0)}_{\rho\pi\pi}}
 \biggr],
\end{eqnarray}
 where
 $$\varepsilon(s) = {-\Pi_{\rho\omega} \over {D_\omega(s)-D_\rho(s)}}, 
 \mbox{~~}
 |g_{V\gamma}| = \Biggl[ {{3m_V^3\Gamma_VB(V \to e^+e^-)} \over
 {4\pi\alpha}} \Biggr]^{1/2}, \mbox{~~}
 |g_{V\pi\pi}| = \Biggl[{{6\pi m^2_V\Gamma_VB(V\to\pi^+\pi^-)} \over
 q^3_\pi(m_V)} \Biggr]^{1/2}.$$
 The superscript $(0)$ denotes the coupling constants of the bare, unmixed
 state. $\Pi_{\rho\omega}$ is the polarization operator of the $\rho-\omega$
 mixing:
\begin{eqnarray}
 \Pi_{\rho\omega}(s) = \mbox{Re}(\Pi_{\rho\omega}(s)) + 
 i\mbox{~}\mbox{Im}(\Pi_{\rho\omega}(s)).
\end{eqnarray}
 The $\mbox{Im}(\Pi_{\rho\omega}(s))$ can be written as
\begin{eqnarray}
 \mbox{Im}(\Pi_{\rho\omega}(s))=\sqrt{s}\biggl\{
 {g^{(0)}_{\rho\pi\pi}g^{(0)}_{\omega\pi\pi}q^3_\pi(s) \over {6\pi s}} +
 {g^{(0)}_{\rho\pi\gamma}g^{(0)}_{\omega\pi\gamma}q^3_{\pi\gamma}(s) +
  g^{(0)}_{\rho\eta\gamma}g^{(0)}_{\omega\eta\gamma}q^3_{\eta\gamma}(s)
  \over 3},
 \biggr\}
\end{eqnarray}
 where
$$
 g_{VP\gamma}=
 \biggl[{3\Gamma_VB(V\to P\gamma)\over q^3_{P\gamma}(m_V)}\biggr]^{1/2}.
$$
 We neglected the contributions to $\mbox{Im}(\Pi_{\rho\omega}(s))$ due to 
 $VP$ intermediate state ($V=\omega,\rho$,$P=\pi,\eta$).
 The $\mbox{Re}(\Pi_{\rho\omega}(s))$ can be represented as
\begin{eqnarray}
 \mbox{Re}(\Pi_{\rho\omega}(s)) = \mbox{Re}(\Pi^\gamma_{\rho\omega}(s)) +
 \mbox{Re}(\Pi^\prime_{\rho\omega}(s)),
\end{eqnarray}
 where
\begin{eqnarray}
 \mbox{Re}(\Pi^\gamma_{\rho\omega}(s))=
 {-4\pi g^{(0)}_{\rho\gamma}g^{(0)}_{\omega\gamma}\over s}
\end{eqnarray}
 represents the one-photon contribution to the 
 $\mbox{Re}(\Pi_{\rho\omega}(s))$. Let us assume that the energy dependence 
 of the $\mbox{Re}(\Pi^\prime_{\rho\omega}(s))$ is negligible, then
 it can be expressed by using the measured branching ratio
\begin{eqnarray}
 B(\omega\to\pi^+\pi^-)={\Gamma_{\rho}(m_\omega)\over\Gamma_\omega}
 \biggl| \varepsilon(m_\omega)+
 {g^{(0)}_{\omega\pi\pi}\over g^{(0)}_{\rho\pi\pi}}
 \biggr|^2
\end{eqnarray}
 as follows
\begin{eqnarray}
 \mbox{Re}(\Pi^\prime_{\rho\omega}) =
 {4\pi g^{(0)}_{\rho\gamma}g^{(0)}_{\omega\gamma}\over m^2_\omega} +
 {g^{(0)}_{\omega\pi\pi}\over g^{(0)}_{\rho\pi\pi}}(m^2_\omega-m^2_\rho) +
 \nonumber \\ 
 + \sqrt{
 {\Gamma_\omega B(\omega\to\pi^+\pi^-)\over\Gamma_{\rho}(m_\omega)}
 \biggl|D_\omega(m_\omega)-D_\rho(m_\omega)\biggr|^2-
 \biggl[{g^{(0)}_{\rho\pi\gamma}g^{(0)}_{\omega\pi\gamma}q^3_{\pi\gamma}
 (m_\omega)+
  g^{(0)}_{\rho\eta\gamma}g^{(0)}_{\omega\eta\gamma}q^3_{\eta\gamma}(m_\omega)
 \over 3}+
 {g^{(0)}_{\omega\pi\pi}\over g^{(0)}_{\rho\pi\pi}}m_\omega\Gamma_\omega
 \biggr]^2}
\end{eqnarray}
 
 Equation (\ref{rhoom}) can be rewritten as follows
\begin{eqnarray}
\label{amprhoom}
 A_{\omega\to\pi^+\pi^-} + A_{\rho\to\pi^+\pi^-} = 
 \sqrt{3\over2} {1\over\alpha}   
 \sum_{V=\omega,\rho}
 { {\Gamma_V m_V^3 \mbox{~} \sqrt[]{m_V\sigma(V\to\pi^+\pi^-)} } 
 \over {D_V(s)} } 
 { f_{V\pi\pi}(s)\over\sqrt[]{q_\pi(m_V)}},
\end{eqnarray}
 where
 $$f_{V\pi\pi}(s) = {r_{V\pi\pi}(s) \over r_{V\pi\pi}(m_V) }, \mbox{~~}$$
 and 
 $$r_{\rho\pi\pi}(s) =
 1-{g^{(0)}_{\gamma\omega}\over g^{(0)}_{\gamma\rho}}\varepsilon(s), 
 \mbox{~~} r_{\omega\pi\pi}(s)=\varepsilon(s)+
 {g^{(0)}_{\omega\pi\pi}\over g^{(0)}_{\rho\pi\pi}}$$
 The theoretical value of the phase $\phi_{\rho\omega}$ can be calculated from
 the above given expressions:
 $\phi_{\rho\omega}=\arg(f_{\omega\pi\pi}(m_\omega))-
 \arg(f_{\rho\pi\pi}(m_\rho))\simeq 101^\circ$.
 The phase $\phi_{\rho\omega}$ almost does not depend on energy. In this
 calculation we assumed that the $\omega\to\pi^+\pi^-$ transition proceeds
 only via the $\rho-\omega$ mixing, that is $g^{(0)}_{\omega\pi\pi}=0$. In
 order to determine the $g^{(0)}_{\rho\pi\pi}$, $g^{(0)}_{\gamma V}$ and
 $g^{(0)}_{VP\gamma}$ coupling constants, the corresponding measured decay 
 widths were used. 

\subsection{Fit to the experimental data}
 The $\rho^\prime$ and $\rho^{\prime\prime}$ parameters were determined from
 the fit to the $e^+e^-\to\pi^+\pi^-$ cross section measured at
 the energy region $\sqrt{s}<2400$ MeV by OLYA and DM2 detectors 
 \cite{olya,dm2}, together with the isovector part of the 
 $e^+e^-\to\pi^+\pi^-$ cross section calculated by assuming
 the CVC hypothesis from the spectral function of the 
 $\tau^-\to\pi^-\pi^0\nu_\tau$ decay measured by CLEO II \cite{cleo2}:
\begin{eqnarray}
 \sigma_{\pi\pi}(m_i)={4(\pi\alpha)^2\over m_i} 
 {B(\tau\to\pi\pi^0\nu_\tau)\over B(\tau\to e \overline{\nu}_e\nu_\tau)}
 {m^8_\tau\over 12\pi |V_{ud}|^2}{1\over S_{EW}}
 {1\over m_i(m^2_\tau-m^2_i)^2(m^2_\tau+2m^2_i)}
 {1\over N} {N_i\over\Delta m_i},
\end{eqnarray}
 where $m_i$ is the central value of the $\pi^-\pi^0$ pair invariant mass for
 the  $i$-th bin, $\Delta m_i$ is the bin width, $N_i$ is the number of
 entries in the $i$-th bin, $N$ is the total number of entries, $|V_{ud}|$ 
 is the CKM matrix element, $S_{EW}=1.0194$ is the radiative correction
 \cite{aleph,cleo2,radtau}. 
 
\begin{figure}
\begin{center}
\epsfig{figure=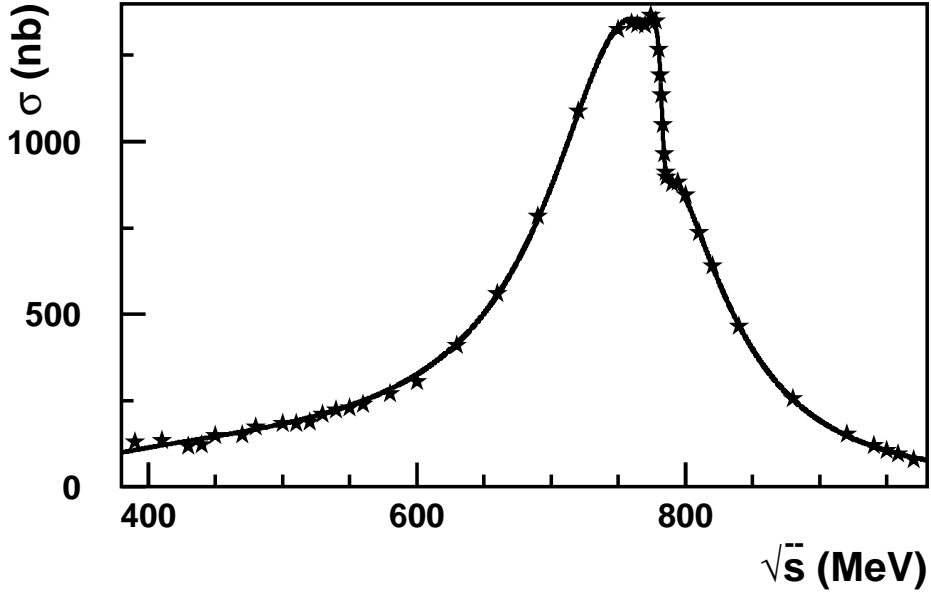,width=15.0cm}
\caption{The $e^+e^-\to\pi^+\pi^-$ cross section.
         Stars are the SND data obtained in this work, curve is the fit
	 result.}
\label{sndfit}
\end{center}
\end{figure}
 
 The obtained $\rho^\prime$ and $\rho^{\prime\prime}$ parameters were used
 in the fitting to the SND data (Table~\ref{tab5}, Fig.\ref{sndfit}).
 The free parameters of the fit were $m_\rho$, $\Gamma_{\rho}$, 
 $\sigma(\rho\to\pi^+\pi^-)$, $\sigma(\omega\to\pi^+\pi^-)$, 
 $\phi_{\rho\omega}$ and $\sigma(\rho^\prime\to\pi^+\pi^-)$.
 The first fit was performed with $\sigma(\rho^{\prime\prime}\to\pi^+\pi^-)$,
 $\rho^\prime$ and $\rho^{\prime\prime}$ masses and widths fixed at the
 values obtained from the fit to the CLEO II and DM2 data. The second and 
 third fits were done without the $\rho^{\prime\prime}$ meson. The 
 $\rho^\prime$ mass
 and width were fixed by using results of the fit to the CLEO II and DM2 data
 (the second variant in the Table~\ref{tab5}) and to the OLYA data (the third
 variant in the
 Table~\ref{tab5}). The values of the $\rho$ and $\omega$ parameters exhibit
 a rather weak model dependence.
 
\begin{table}
\caption{Fit results. The column number $N$ corresponds to the different
         variants of choice of the  $\rho^\prime$ and $\rho^{\prime\prime}$ 
         parameters.}
\label{tab5}
\begin{center}
\begin{tabular}[t]{llll}
 N  &1&2&3 \\ \hline
$m_\rho,$MeV&774.9$\pm$0.4&774.9$\pm$0.4&774.9$\pm$0.4 \\
$\Gamma_\rho,$MeV&146.2$\pm$0.8&146.4$\pm$0.8&146.3$\pm$0.8 \\
$\sigma(\rho\to\pi^+\pi^-),$ nb&1222$\pm$7&1218$\pm$7&1219$\pm$7 \\
$\sigma(\omega\to\pi^+\pi^-)$,nb&30.2$\pm$1.4&30.3$\pm$1.4&30.3$\pm$1.4 \\
$\phi_{\rho\omega}$, degree&113.6$\pm$1.3&113.4$\pm$1.3&113.5$\pm$1.3 \\
$m_{\rho^\prime}$, MeV&1403&1403&1360 \\
$\Gamma_{\rho^\prime}$, MeV&455&455&430 \\
$\sigma(\rho^\prime\to\pi^+\pi^-)$,nb&3.8$\pm$0.3&1.8$\pm$0.2&1.9$\pm$0.2 \\
$m_{\rho^{\prime\prime}}$, MeV&1756&&\\
$\Gamma_{\rho^{\prime\prime}}$, MeV&245&&\\
$\sigma(\rho^{\prime\prime}\to\pi^+\pi^-)$, nb&1.7&&\\
$\chi^2/N_{df}$&50.2/39&48.8/39&49.4/39 \\ \hline
\end{tabular} 
\end{center}
\end{table}

\section{Discussion.}

\begin{figure}
\begin{center}
\epsfig{figure=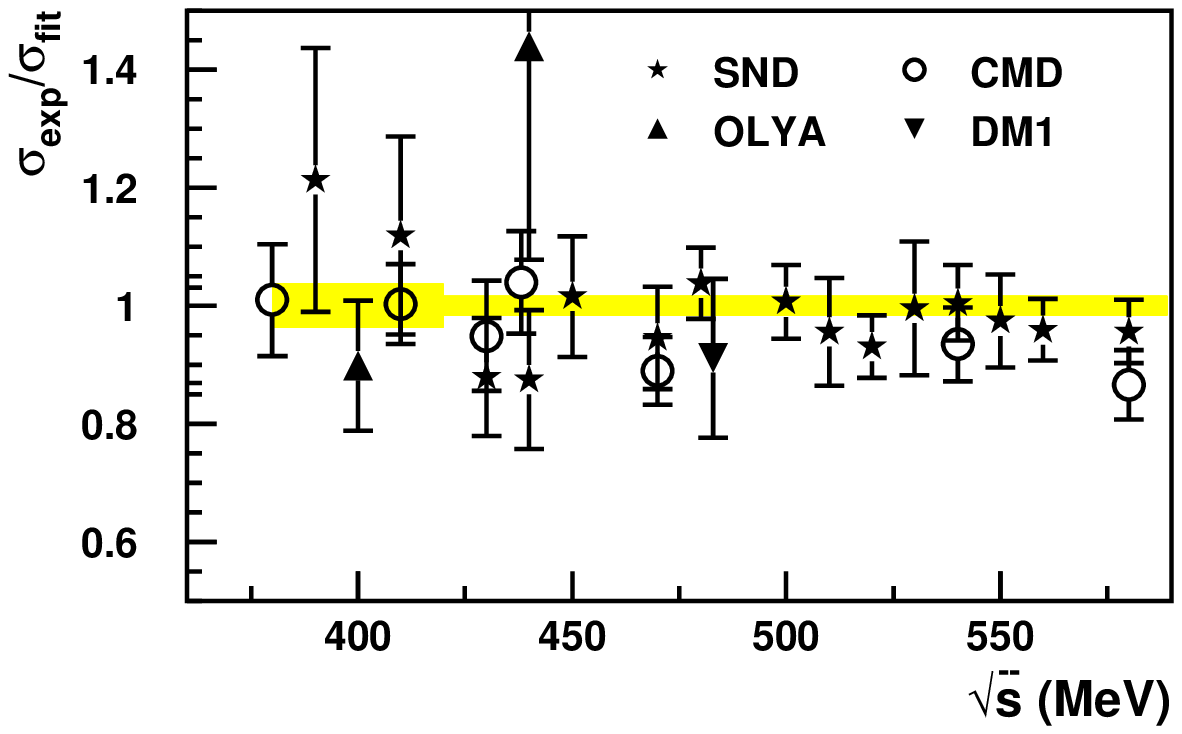,width=15.0cm}
\caption{The ratio of the  $e^+e^-\to\pi^+\pi^-$ cross section obtained
        in different experiments to the fit curve (Fig.\ref{sndfit}).
	The shaded area shows the systematic error of the SND measurements.
        The SND (this work), CMD,OLYA and DM1 \cite{quen,olya} results are
        presented.}
\label{otn1}
\epsfig{figure=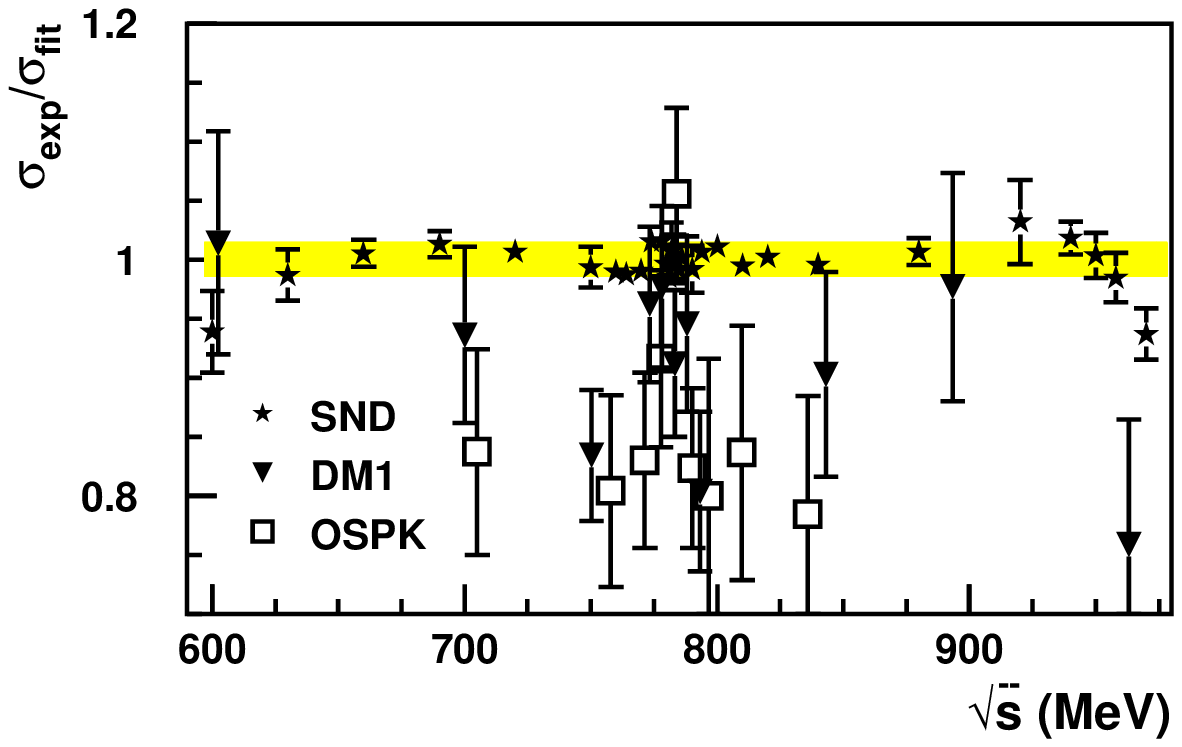,width=15.0cm}
\caption{The ratio of the  $e^+e^-\to\pi^+\pi^-$ cross section obtained
        in different experiments to the fit curve (Fig.\ref{sndfit}).
	The shaded area shows the systematic error of the SND measurements.
        The SND (this work), DM1, OSPK \cite{quen,bena} results are 
	presented.} 
\label{otn2}
\end{center}
\end{figure}
\begin{figure}
\begin{center}
\epsfig{figure=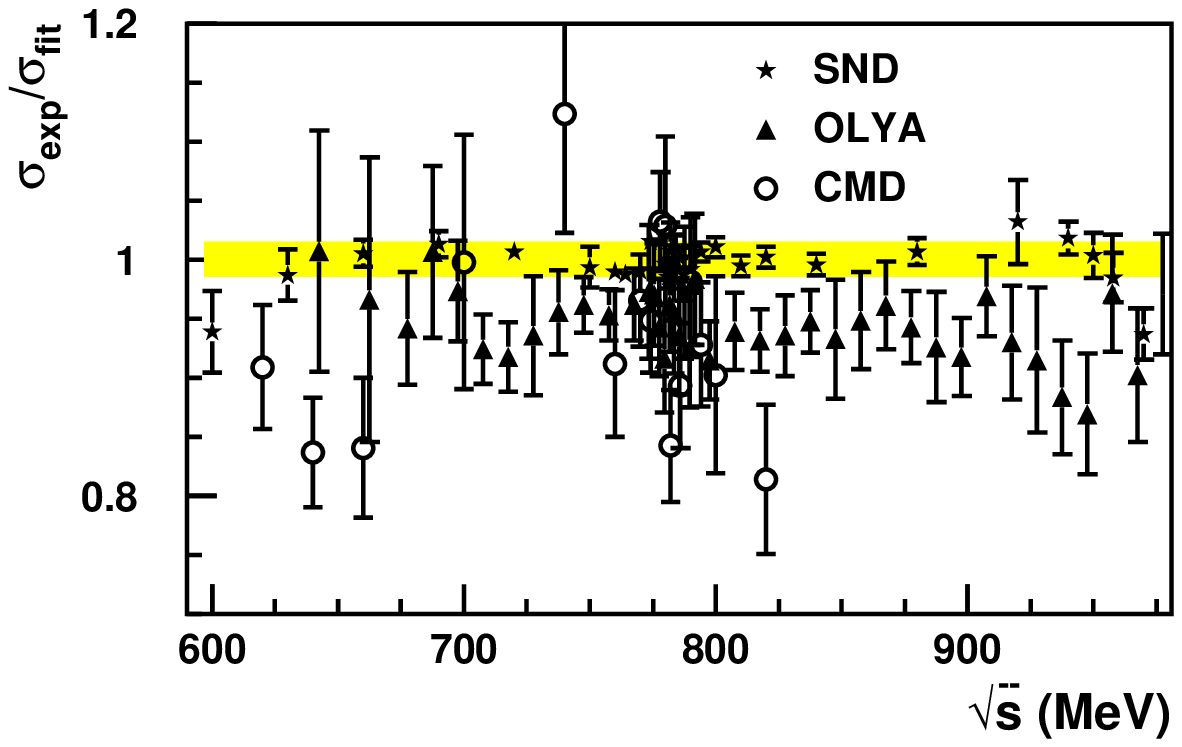,width=15.0cm}
\caption{The ratio of the  $e^+e^-\to\pi^+\pi^-$ cross section obtained
        in different experiments to the fit curve (Fig.\ref{sndfit}).
	The shaded area shows the systematic error of the SND measurements.
        The SND (this work), OLYA and CMD \cite{olya} results are presented.}
\label{otn3}
\epsfig{figure=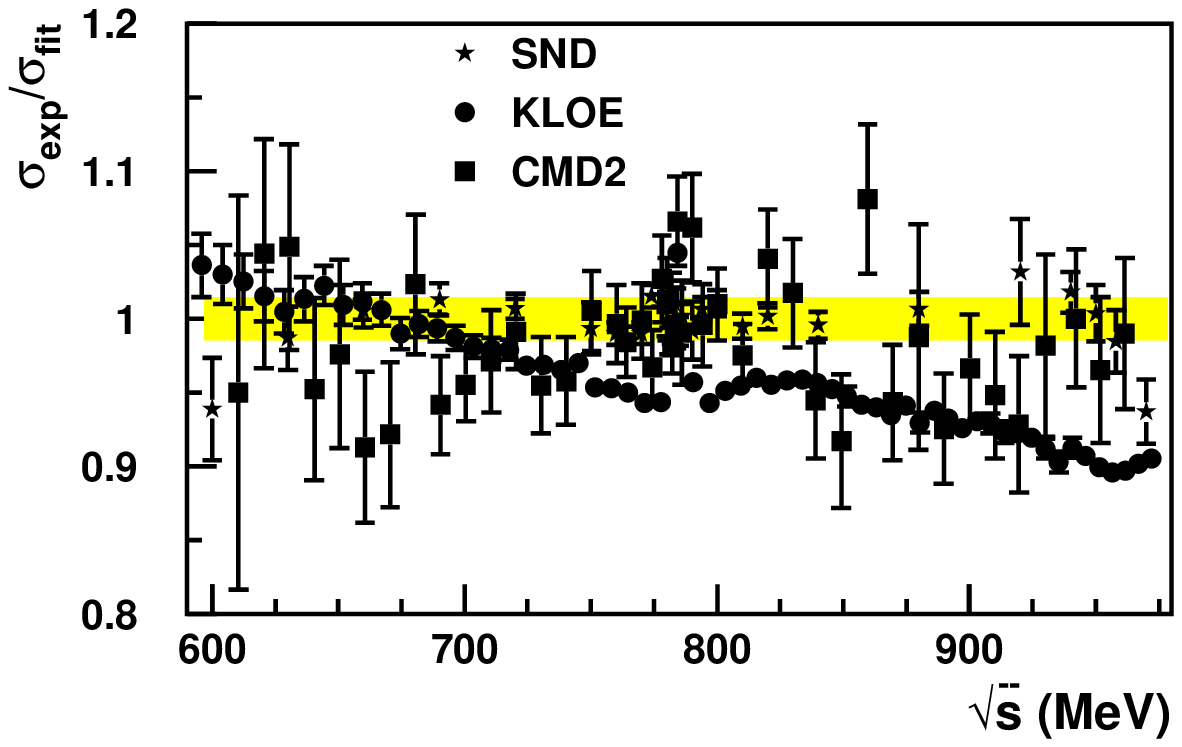,width=15.0cm}
\caption{The ratio of the  $e^+e^-\to\pi^+\pi^-$ cross section obtained
        in different experiments to the fit curve (Fig.\ref{sndfit}).
	The shaded area shows the systematic error of the SND measurements.
        The SND (this work), CMD2 and KLOE \cite{kloe,kmd2} results are
        presented.} 
\label{otn4}
\end{center}
\end{figure}
 The comparison of the  $e^+e^-\to\pi^+\pi^-$ cross section obtained in SND
 experiment with other results \cite{bena,quen,olya,kmd2,kloe} is
 shown in Fig.\ref{otn1},\ref{otn2},\ref{otn3}~and~\ref{otn4}. In the
 energy region $\sqrt{s}<600$ MeV all experimental data are in agreement
 (Fig.\ref{otn1}). Above 600 MeV the OSPK(ORSAY-ACO)\cite{bena} and DM1 
 \cite{quen} points lay about 10 \% lower than the SND ones (Fig.\ref{otn2}).
 The SND cross section exceeds the OLYA and CMD measurements \cite{olya}
 by $6\pm 1$ \% in this energy region (Fig.\ref{otn3}).
 The systematic error of OLYA measurement is 4 \% and the OLYA data agree with
 the SND result. The systematic uncertainty of CMD result is 2 \%, so the 
 difference between the SND and CMD results is about 2.5 of joint systematic 
 error. At the same time
 the SND and CMD data below 600 MeV agree well (Fig.\ref{otn1}). The 
 average deviation between CMD2 \cite{kmd2} and SND data is
 $1.4 \pm 0.5$ \%, the systematic inaccuracies of these measurements are 
 0.6 \% and 1.3 \% respectively. In the KLOE experiment at $\phi$-factory
 DAF$\Phi$NE the form factor $|F_\pi(s)|^2$ was measured by using 
 ``radiative return'' method with systematic error of 0.9 \% \cite{kloe}. 
 In Ref.\cite{kloe} the bare form factor is listed. So in order to compare
 the KLOE result with the SND one, the form factor was appropriately dressed 
 by us. The results of this
 comparison are shown in Fig.\ref{otn4}. The KLOE measurement is in conflict
 with the SND result as well as with the CMD2 one.

\begin{figure}
\begin{center}
\epsfig{figure=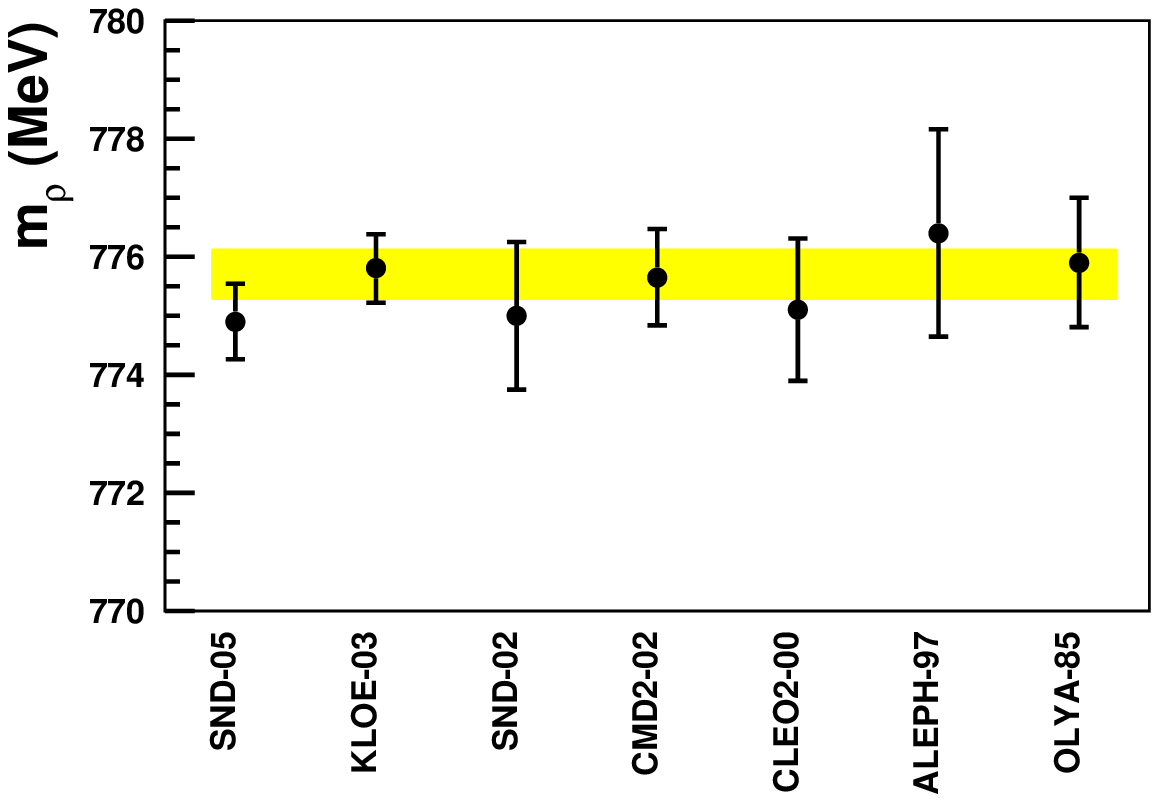,width=15.0cm}
\caption{The $\rho$-meson mass $m_\rho$ measured in this work (SND-05)
         and in Ref.\cite{kloe3pi,dplphi98,kmd2,cleo2,aleph,olya}. 
	 The shaded area shows the average of the previous results.}
\label{massa}
\epsfig{figure=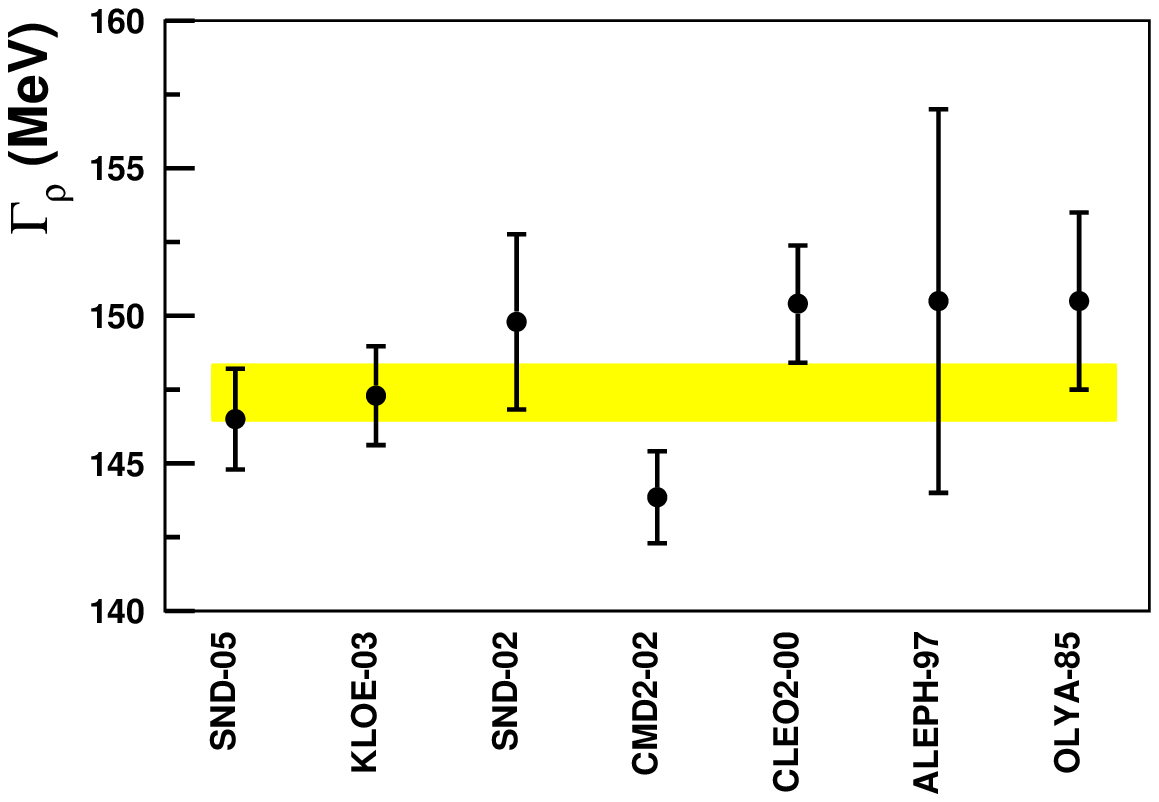,width=15.0cm}
\caption{The $\rho$ meson width $\Gamma_\rho$ measured in this work (SND-05)
         and in Ref.\cite{kloe3pi,dplphi98,kmd2,cleo2,aleph,olya}. 
         The shaded area shows the average of the previous results.}
\label{shira}
\end{center}
\end{figure}

 The $\rho$-meson parameters $m_\rho$, $\Gamma_\rho$,
 $\sigma(\rho\to\pi^+\pi^-)$ were determined from study of the
 $e^+e^-\to\pi^+\pi^-$ cross section. The $\rho$ meson mass and width were
 found to be
 $$ m_\rho = 774.9 \pm 0.04 \pm 0.05 \mbox{~~MeV}, $$
 $$\Gamma_\rho = 146.5 \pm 0.8 \pm 1.5 \mbox{~~MeV}.$$
 The systematic errors is related to the accuracy of the collider energy
 determination, to the model uncertainty and to the error of the cross 
 section determination. The $\rho$-meson parameters were studied in other 
 $e^+e^-$ experiments by 
 using the processes $e^+e^-\to\pi^+\pi^-$ \cite{kmd2,olya},
 $e^+e^-\to\rho\pi\to\pi^+\pi^-\pi^0$ \cite{kloe3pi,dplphi98} and the 
 $\tau^-\to\pi^-\pi^0\nu_\tau$ decay \cite{cleo2,aleph}. The SND results 
 are in agreement with these measurements as is shown in
 Fig.\ref{massa} and \ref{shira}. 

\begin{figure}
\begin{center}
\epsfig{figure=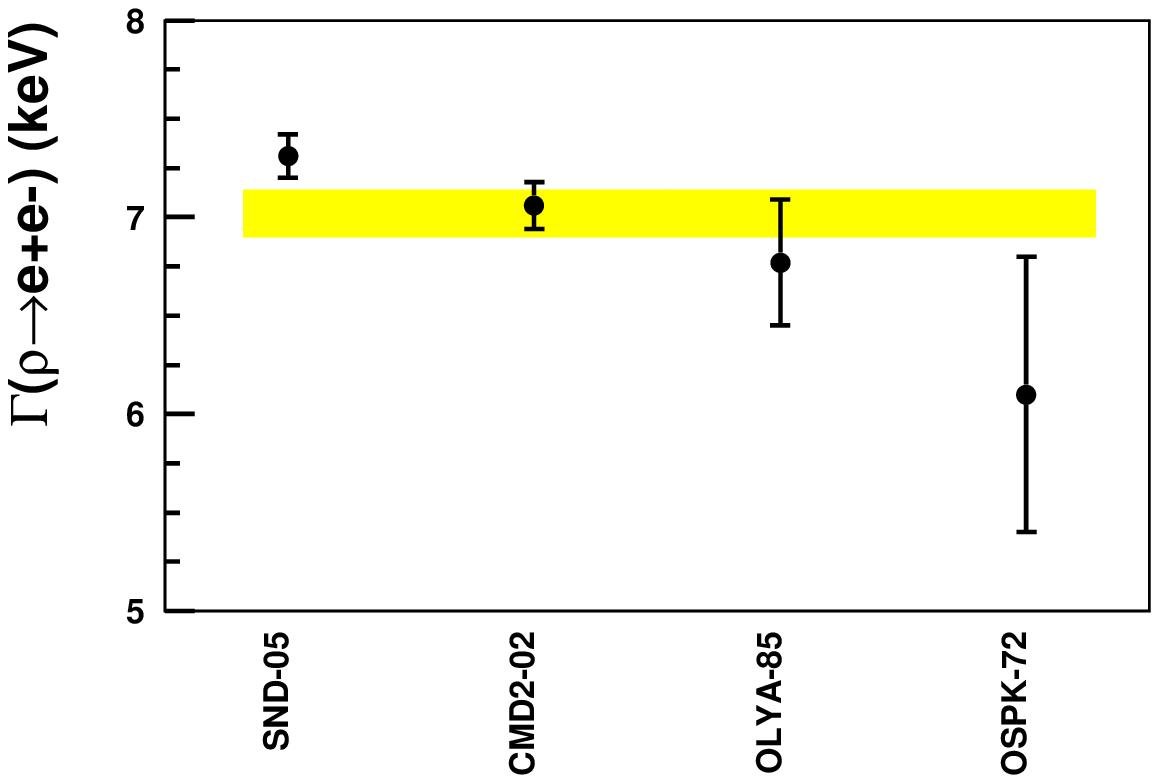,width=15.0cm}
\caption{The value of $\Gamma(\rho\to e^+e^-)$ obtained in this work (SND-05)
         and in Ref.\cite{kmd2,olya,bena}. 
	 The shaded area shows the world average value \cite{pdg}.}
\label{poee}
\epsfig{figure=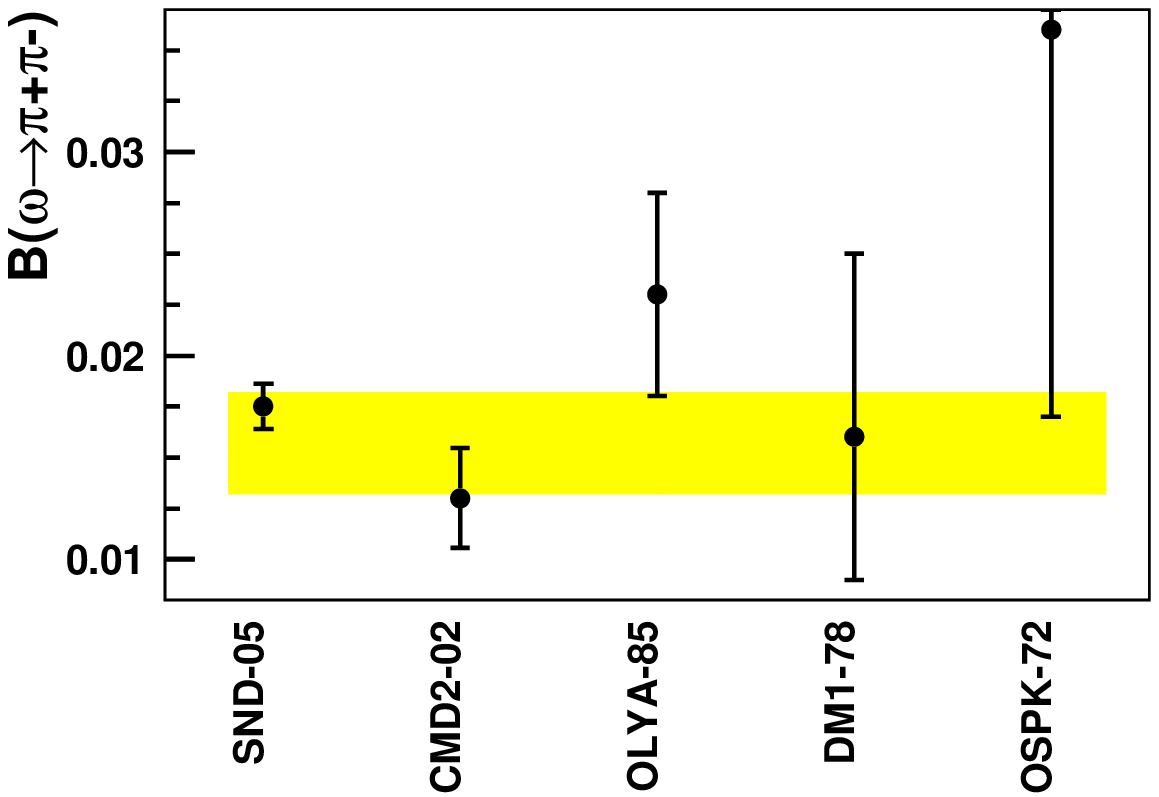,width=15.0cm}
\caption{The value of $B(\omega\to\pi^+\pi^-)$ obtained in this work (SND-05)
          and in Ref.\cite{kmd2,olya,quen,bena}.  
	  The shaded area shows the world average value \cite{pdg}.}
\label{om2p}
\end{center}
\end{figure}

 The parameter $\sigma(\rho\to\pi^+\pi^-)$ was found to be
 $$\sigma(\rho\to\pi^+\pi^-) = 1220 \pm 7 \pm 16 \mbox{~~nb},$$
 which corresponds to
 $$B(\rho\to e^+e^-)\times B(\rho\to\pi^+\pi^-)=
   (4.991\pm 0.028\pm0.066)\times 10^{-5},$$
 $$\Gamma(\rho\to e^+e^-)=7.31\pm 0.021\pm0.11 \mbox{~~keV}.$$
 The systematic error includes the systematic uncertainties in the cross
 section measurement and the model dependence. 
 A comparison of the $\Gamma(\rho\to e^+e^-)$  obtained in this work with 
 other experimental results \cite{kmd2,olya,bena} and with the PDG world
 average \cite{pdg} is shown in Fig.\ref{poee}. The SND result exceeds all
 previous measurements. It differs by about
 1.5 standard deviations from the CMD2 measurement \cite{kmd2} and by 2 
 standard deviations 
 from the PDG world average \cite{pdg}. The difference of the $\rho$-meson
 leptonic widths obtained by SND and CMD2 should be 
 attributed mainly to the difference in the total widths of the $\rho$-meson
 rather then to the difference in the cross section values. The value
 $\sigma(\rho\to\pi^+\pi^-)=1198$ nb, which can be obtained by using CMD2 
 cross section data 
 reported in Ref.\cite{kmd2}, agrees with the SND result within the 
 measurements errors.

 The parameter $\sigma(\omega\to\pi^+\pi^-)$ was found to be
 $$\sigma(\omega\to\pi^+\pi^-) = 29.9 \pm 1.2 \pm 1.0 \mbox{~~nb},$$
 which corresponds to
 $$B(\omega\to e^+e^-)\times B(\omega\to\pi^+\pi^-)=
   (1.247\pm 0.062\pm0.042)\times 10^{-6}.$$
 The systematic error is related to the model dependence, to the error of the
 cross section determination and to the accuracy of the collider energy 
 determination.
 In the previous studies of the  $e^+e^-\to\pi^+\pi^-$ reaction the
 relative probability of the $\omega\to\pi^+\pi^-$ decay was also reported.
 The comparison of $B(\omega\to\pi^+\pi^-)=0.0175\pm 0.0011$ obtained by using
 the SND data and the PDG value of the $\omega\to e^+e^-$ decay width 
 \cite{pdg} with the results of other experiments is shown in Fig.\ref{om2p}.
 The SND result is the most precise.
 
\begin{figure}
\epsfig{figure=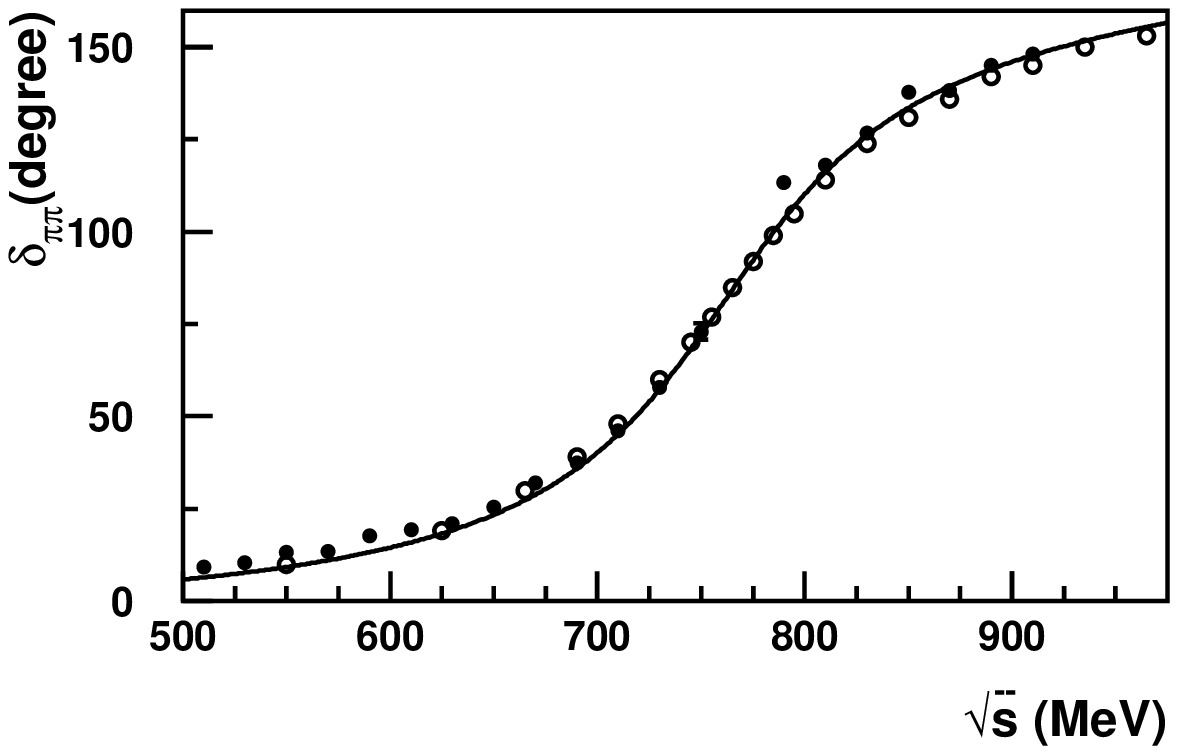,width=15.0cm}
\caption{The $\pi\pi$ scattering phase in the P-wave. Dots and circles are
         results of the phase measurements in Ref. \cite{pwa1,pwa2} by using 
         the reaction $\pi N\to\pi\pi N$ . The curve is the phase
	 of the amplitude $A_{\rho\to\pi\pi}+A_{\rho\to\pi^+\pi^-}$ obtained
	 from the fit to the SND data presented in this work.}
\label{ppfaz}
\epsfig{figure=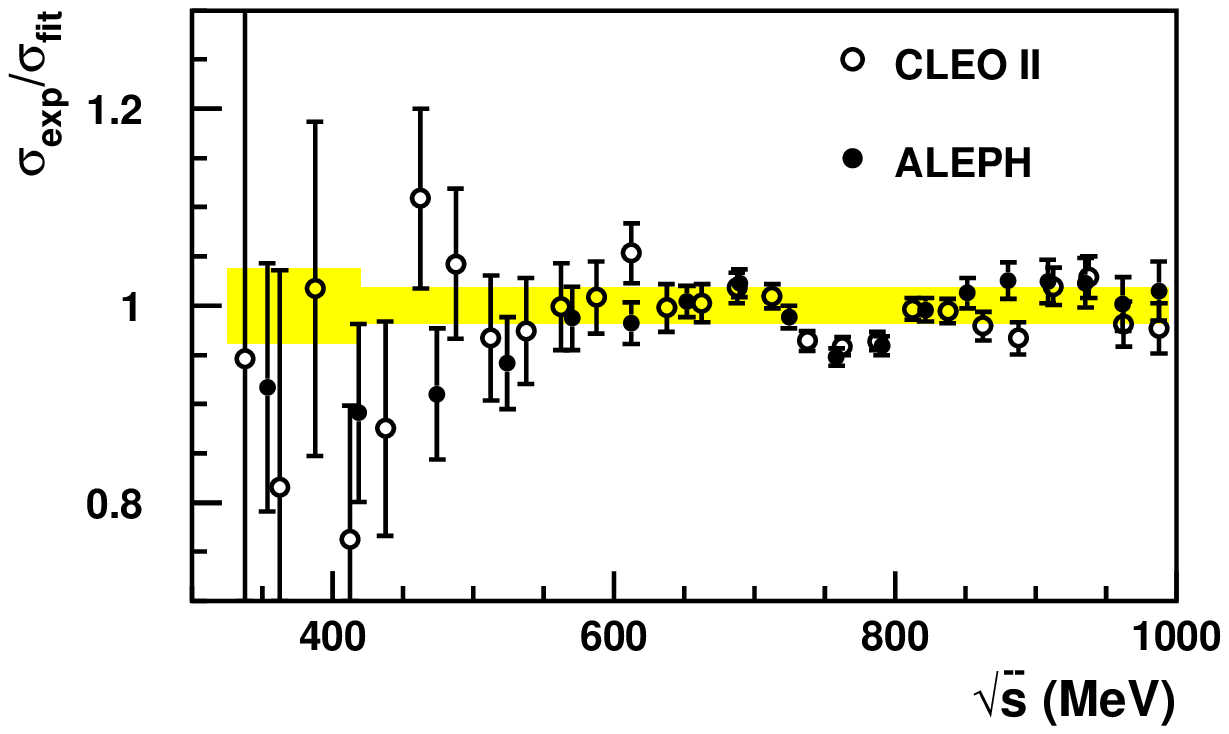,width=15.0cm}
\caption{The ratio of the $e^+e^-\to\pi^+\pi^-$ cross section calculated
         from the  $\tau^-\to\pi^-\pi^0\nu_{\tau}$ decay spectral function
	 measured in Ref.\cite{cleo2,aleph} to the isovector part of the
	 $e^+e^-\to\pi^+\pi^-$ cross section measured in this work
	 The shaded area shows the joint systematic error.}
\label{tau1}
\end{figure}

 The phase $\phi_{\rho\omega}$ was found to be
 $$\phi_{\rho\omega}=113.5 \pm 1.3 \pm 1.7 \mbox{~~degree}.$$
 This value differs by  6 standard deviations from $101^\circ$ expected
 under assumption that the $\omega\to\pi^+\pi^-$ transition proceeds through
 the $\rho-\omega$ mixing mechanism. If instead of the phase 
 $\phi_{\rho\omega}$,
 the ratio ${g^{(0)}_{\omega\pi\pi}/ g^{(0)}_{\rho\pi\pi}}$ is the free 
 parameter of the fit it follows that
 $${g^{(0)}_{\omega\pi\pi}\over g^{(0)}_{\rho\pi\pi}}=0.11\pm 0.01.$$
 This ratio corresponds to the too large direct transition width
 $\Gamma^{(0)}(\omega\to\pi^+\pi^-)=1.82\pm 0.33$ MeV, while the natural 
 expectation is
 $\Gamma^{(0)}(\omega\to\pi^+\pi^-)\approx\alpha^2\Gamma_\rho\approx 8$ keV.
 Let us note, that the analysis of the OLYA and CMD2 data \cite{kmd2,olya}
 give the similar values of the $\phi_{\rho\omega}$ phase. This result can 
 point out that the considerable direct transition $\omega\to\pi^+\pi^-$  
 exists. On the other hand this discrepancy can be attributed also to 
 inadequacies of the applied theoretical model.

 The comparison of the phase
 $\arg(A_{\rho\to\pi^+\pi^-}+A_{\rho^\prime\to\pi^+\pi^-})$ with the
 $\pi\pi$ scattering phase in the P-wave \cite{pwa1,pwa2} is shown
 in Fig.\ref{ppfaz}. These phases must be equal in the purely elastic
 scattering region. The agreement is satisfactory, in any case in the 
 energy region $\sqrt{s}\approx m_\rho$ no essential difference is
 observed. 

 The comparison of the $e^+e^-\to\pi^+\pi^-$ cross section, obtained
 under the CVC hypothesis from the $\tau$ spectral function of the
 $\tau^-\to\pi^-\pi^0\nu_{\tau}$ decay \cite{cleo2,aleph} with isovector 
 part of the cross section measured in this work is shown in Fig.\ref{tau1}.
 The cross section obtained by SND was undressed from the vacuum polarization
 and the contribution 
 from the $\omega\to\pi^+\pi^-$ decay was excluded. The cross section
 calculated from the  $\tau$ spectral function was multiplied by the
 coefficient which takes into account the difference of the $\pi^\pm$ and
 $\pi^0$ masses:
 $$
 \delta = \biggl({q_\pi(s) \over q_{\pi^\pm}(s)}\biggr)^3
 {|A_{\pi^+\pi^-}(s)|^2 \over |A_{\pi^0\pi^\pm}(s)|^2}, \mbox{~~~}
 q_{\pi^\pm}(s) = {1 \over 2\sqrt{s}}
 \bigl[(s-(m_{\pi^0}+m_{\pi^\pm})^2)(s-(m_{\pi^0}-m_{\pi^\pm})^2)\bigr]^{1/2}.
 $$
 The average deviation of the SND and $\tau$ data is about 1.5 \%.
 For almost all energy points this deviation is within the joint systematic 
 error $\simeq 1.6\%$. The 10\% difference between $e^+e^-$ and $\tau$ data
 at $\sqrt{s}>800$ MeV, which was claimed in Ref.\cite{eetau}, is absent.

 Using the  $\sigma^{pol}_{\pi\pi}(s)$ cross section (Table~\ref{tab1}),
 the contribution to the anomalous magnetic moment of the muon, due to the 
 $\pi^+\pi^-(\gamma)$ intermediate state in the vacuum polarization, was 
 calculated via the dispersion integral:
$$ a_\mu(\pi\pi, 390\mbox{MeV}\le\sqrt[]{s}\le 970\mbox{MeV})=
\biggl({\alpha m_\mu \over 3\pi} \biggr)^2 \int^{s_{max}}_{s_{min}} 
{R(s)K(s) \over s^2} ds,
$$
 where $s_{max}=970$ MeV, $s_{min}=390 MeV$, $K(s)$ is the known kernel and
$$
 R(s) = {\sigma^{pol}_{\pi\pi} \over \sigma(e^+e^-\to\mu^+\mu^-)}, 
 \mbox{~~} \sigma(e^+e^-\to\mu^+\mu^-) = {4 \pi \alpha^2 \over 3 s}.
$$ 

 The integral was evaluated by using the trapezoidal rule. To take into 
 account the numerical integration errors, the correction method suggested 
 in Ref.\cite{aki} was applied. As a result we obtained
 $$a_\mu(\pi\pi, 390\mbox{MeV}\le\sqrt[]{s}\le 970\mbox{MeV}) = 
 (488.7 \pm 2.6 \pm 6.6) \times 10^{-10}.$$
 This is about 70 \% of the total hadronic contribution to the anomalous
 magnetic moment of the muon $(g-2)/2$.

 If the integration is performed  for the energy region corresponding to the
 CMD2 measurements \cite{kmd2}, then the result is 
 $a_\mu(\pi\pi)=(385.6\pm 5.2) \times 10^{-10}$, which is
 1.8 \% (1 standard deviation) higher than the CMD2 result:
 $a_\mu(\pi\pi)=(378,6\pm 3.5) \times 10^{-10}$. So no considerable 
 difference between the SND and CMD2 results is observed.

\section{Conclusion}
 The cross section of the process $e^+e^-\to \pi^+\pi^-$  was measured in
 the SND experiment at the VEPP-2M collider in the energy region
 $390<\sqrt[]{s}<980$ MeV with accuracy  1.3 \% at $\sqrt{s}\ge 420$ MeV and
 3.4 \% at $\sqrt{s}<420$ MeV. The measured cross section was analyzed in the
 framework of the generalized vector meson dominance model.
 The following $\rho$-meson parameters were obtained: 
 $m_\rho=774.9\pm 0.4\pm 0.5$ MeV, 
 $\Gamma_\rho=146.5 \pm 0.8 \pm 1.5$ MeV and
 $\sigma(\rho\to\pi^+\pi^-)=1220\pm 7\pm 16$ nb. The 
 parameters of the  $G$-parity suppressed process 
 $e^+e^-\to\omega\to\pi^+\pi^-$  
 were measured with high precision. The measured value 
 $\sigma(\omega\to\pi^+\pi^-)=29.9\pm 1.4\pm 1.0$ nb corresponds to
 the relative probability $B(\omega\to\pi^+\pi^-) = 1.75 \pm 0.11 \%$. 
 The relative interference phase between the $\rho$ and $\omega$ mesons was
 found to be $\phi_{\rho\omega} = 113.5\pm 1.3\pm 1.7$ degree. This
 result is in conflict with the naive expectation from the $\rho-\omega$ mixing
 $\phi_{\rho\omega} =101^\circ$.
 The SND result agrees with the cross section calculated from the $\tau$ 
 spectral function data within the accuracy of the measurements.
 Using measured cross section, the contribution to the anomalous magnetic
 moment of the muon due to the $\pi^+\pi^-(\gamma)$ intermediate state  in the 
 vacuum polarization was calculated: 
 $a_\mu(\pi\pi, 390\mbox{MeV}\le\sqrt[]{s}\le 970\mbox{MeV}) =
 (488.7 \pm 2.6 \pm 6.6) \times 10^{-10}.$

\acknowledgments
 The authors are grateful to N.N.Achasov for useful discussions.
 The work is supported in part by grants Sci.School-1335.2003.2,
 RFBR 04-02-16181-a, 04-02-16184-a, 05-02-16250-a.


\begin{thebibliography}{99}
\bibitem{bnl1}
 H.N. Brown et al., Phys. Rev. Lett. {\bf 86}, 2227 (2001)
\bibitem{bnl2}
 G.W. Bennet et al., Phys. Rev. Lett. {\bf 92}, 161802 (2004)
\bibitem{aleph}
 R. Barate et al., Z. Phys. C {\bf 76}, 15 (1997)
\bibitem{opal}
 K. Ackerstaff et al., Eur. Phys. J. C {\bf 7}, 571, (1999)
\bibitem{cleo2}
 S. Anderson et al., Phys. Rev. D {\bf 61}, 112002, (2000)
\bibitem{augu}
 J.E. Augustin et al., Phys. Rev. Lett. {\bf 20}, 126, (1968); \\
 J.E. Augustin et al., Nuovo Cim. Lett. {\bf 2}, 214, (1969); \\
 J.E. Augustin et al., Phys. Lett. B{\bf 28}, 508, (1969)
\bibitem{ausl}
 V.L. Auslender et al., Phys. Lett. B{\bf 25}, 433, (1967); \\
 V.L. Auslender et al., Yad. Fiz. {\bf 9}, 114, (1969) [Sov. J. Nucl. Phys.
 {\bf 9}, 69, 1969]
\bibitem{bena} 
 D. Benaksas et al., Phys. Lett. B{\bf 39}, 289, (1972)
\bibitem{quen}
 A. Quenzer et al., Phys. Lett. B{\bf 76}, 512, (1978)
\bibitem{vas1}
 I.B. Vasserman et al., Yad. Fiz. {\bf 28}, 968, (1978)
\bibitem{buki}
 A.D. Bukin et al., Phys. Lett. B{\bf 73}, 226, (1978)
\bibitem{vas2}
 I.B. Vasserman et al., Yad. Fiz. {\bf 30}, 999, (1979)
\bibitem{vas3}
 I.B. Vasserman et al., Yad. Fiz. {\bf 33}, 709, (1981) [Sov. J. Nucl. Phys.
 {\bf 33}, 368, 1981]
\bibitem{kur1}
 L.M. Kurdadze et al., JETP Lett. {\bf 37}, 733, 1983 [Pisma Zh. Eksp. Teor.
 Fiz. {\bf 37}, 613, 1983]
\bibitem{kur2}
 L.M. Kurdadze et al., Yad. Fiz. {\bf 40}, 451, (1984) [Sov. J. Nucl. Phys.
 {\bf 40}, 286, (1984)]
\bibitem{spec}
 S.R. Amendolia et al., Phys. Lett. B{\bf 138}, 454, (1984)
\bibitem{olya}
 L.M. Barkov, et al., Nucl. Phys. B{\bf 256}, 365, (1985)
\bibitem{kmd2}
 R.R. Akhmetshin et al, Phys. Lett. B{\bf 527}, 161, (2002); \\
 R.R. Akhmetshin et al, Phys. Lett. B{\bf 578}, 285, (2004)
\bibitem{kloe}
 A. Aloisio et al., Phys. Lett. B{\bf 606}, 12, (2005)
\bibitem{sndnim}
 M.N. Achasov et al., Nucl. Instr. and Meth. A {\bf 449}, 125 (2000)
\bibitem{vepp2}
 A.N. Skrinsky, in Proc. of Workshop on physics and detectors for
 DA$\Phi$NE, Frascati, Italy, April 4-7, 1995, p.3 
\bibitem{unimod}
 A.D. Bukin et al., in Proc. of Workshop on Detector and Event Simulation in
 High Energy Physics, p.79-85, 8-12 April 1991, NIKHEF, Amsterdam, Netherlands
\bibitem{union}
 A.D. Bukin, N.A. Grozina, Comp. Phys. Comm. {\bf 78}, 287 (1994)
\bibitem{umnuc1}
 K. Hanssgen, S.Ritter, Comp. Phys. Comm. {\bf 31}, 411 (1984); \\
 K. Hanssgen, J.Ranft, Comp. Phys. Comm. {\bf 39}, 37 (1986); \\
 K. Hanssgen, J.Ranft, Comp. Phys. Comm. {\bf 39}, 53 (1986); \\
 K. Hanssgen, Nucl. Sc. and Eng. {\bf 95}, 137 (1987)
\bibitem{umnuc2}
 A.D. Bukin, et al., Report No Budker INP 92-93, Novosibirsk, 1992
\bibitem{berkl}
 F.A. Berends and R. Kleiss, Nucl. Phys. B {\bf 228}, 537 (1983)
\bibitem{arbuzqed}
 A.B. Arbuzov et al., JHEP 10, 001 (1997)
\bibitem{arbuzhad}
 A.B. Arbuzov et al., JHEP 10, 006 (1997)
\bibitem{neural1}
 Bruce H. Denby, Comput.Phys.Commun. {\bf 49}, 429 (1988)
\bibitem{phi98}
 M.N. Achasov et al., Phys. Rev. D {\bf 63}, 072002 (2001)
\bibitem{pi3omeg}
 M.N. Achasov et al., Phys. Rev. D {\bf 68}, 052006 (2003)
\bibitem{dplphi98}
 M.N. Achasov et al., Phys. Rev. D {\bf 65}, 032002 (2002)
\bibitem{bhwide}
 S. Jadach, W. Placzek, B.F.L. Ward, Phys. Lett. B {\bf 390}, 298 (1997)
\bibitem{fedor}
 A.V. Bogdan, et al., Report No Budker INP 2005-33, Novosibirsk, 2005
\bibitem{shw}
 J.Schwinger, Particles, Sources and Fields, vol.II, Addison-Wesley Publishing 
 Company Advanced Book Program Reading, Massachusetts, 1973 \\
 M. Deers, K. Hikasa, Phys. Lett. B {\bf 252}, 127 (1990) \\
 K. Melnikov, Int. J. Mod. Phys. A {\bf 16}, 4591 (2001) \\
 A. Hoefer, J. Gluza, F. Jegerlehner, Eur. Phys. J. C {\bf 24}, 51 (2002)
\bibitem{pi0gam}
 M.N. Achasov et al., Phys. Lett. B {\bf 559}, 171 (2003)
\bibitem{thrhoom}
 N.N. Achasov, A.A. Kozhevnikov, and G.N. Shestakov, Phys. Lett. {\bf 50B},
 448 (1974) .
 N.N. Achasov, N.M. Budnev, A.A. Kozhevnikov, and G.N. Shestakov,
 Yad. Fiz. 23, 610 (1976) [Sov. J. Nucl. Phys. 23, 320 (1976)];
 N.N. Achasov and G.N. Shestakov, Fiz. Elem. Chastits. At. Yadra 9, 48 (1978)
\bibitem{akozi}
 N.N. Achasov and A.A. Kozhevnikov, Yad. Fiz. 55, 809 (1992)
 [Sov. J. Nucl. Phys. 55, 449 (1992)];
 Int. J. Mod. Phys.  A 7, 4825 (1992).  
\bibitem{dm2}
 D.Bisello et al., Phys. Lett. B {\bf 220}, 321 (1989)
\bibitem{radtau}
 E. Braaten, S. Narison and A. Pich, Nucl. Phys. B {\bf 373}, 581 (1992)
\bibitem{kloe3pi}
 A. Aloisio et al., Phys. Lett. B{\bf 561}, 55, (2003)
\bibitem{pdg}
 S. Eidelman et al., Phys. Lett. B {\bf 592}, 1 (2004)
\bibitem{pwa1}
 P. Estabrooks and A.D. Martin, Nucl. Phys. B {\bf 79}, 301 (1974)
\bibitem{pwa2}
 S.D. Protopopescu et al., Phys. Rev. D {\bf 7}, 1279 (1972)
\bibitem{eetau}
  M. Davier et al., Eur. Phys. J. C {\bf27}, 497 (2003) \\
  M. Davier et al., Eur. Phys. J. C {\bf31}, 503 (2003) 
\bibitem{aki}
 N.N. Achasov, A.V.Kiselev, Pisma Zh. Eksp. Teor. Fiz. {\bf75}, 643 (2002)
 [JETP Lett. {\bf 75}, 527 (2002)] 
\end{thebibliography}
\end{document}